\documentclass[11pt,a4paper]{article}

\pdfoutput=1
\usepackage{jheppub}
\usepackage{graphicx}
\usepackage{comment}
\usepackage[normalem]{ulem} % either use this (simple) or
\usepackage{graphics}
\usepackage{feynmp}
\usepackage{epsf}
\usepackage{color}
\usepackage{amsmath}
\usepackage{amssymb}
\usepackage{latexsym}
\usepackage{slashed}
\usepackage{float}
\usepackage{cases}
\usepackage{multirow}

\newcommand{\al}[1]{\begin{align}#1\end{align}}

\newcommand{\bp}{\begin{pmatrix}}
\newcommand{\ep}{\end{pmatrix}}
\newcommand{\bb}{\begin{bmatrix}}
\newcommand{\eb}{\end{bmatrix}}

\newcommand{\paren}[1]{\left(#1\right)}

\newcommand{\beq}{\begin{equation}}
\newcommand{\eeq}{\end{equation}}
\newcommand{\bea}{\begin{eqnarray}}
\newcommand{\eea}{\end{eqnarray}}

\newcommand{\pal}{\partial}

\newcommand{\fulltoday}{\number\day\space \ifcase\month\or
    January\or February\or March\or April\or May\or June\or
    July\or August\or September\or October\or November\or December\fi
    \space\number\year}
\def\rinv{R^{-1}}

\def\rinv{R{^{-1}}}

\def\alphau1{\alpha_1}
\def\alphasu2{\alpha_2}

\title{Non-minimal Universal Extra Dimensions: The strongly interacting
sector at the Large Hadron Collider} 

\author[a]{AseshKrishna Datta}
\author[b]{Kenji Nishiwaki}
\author[b]{Saurabh Niyogi}

\affiliation[a]{Harish-Chandra Research Institute, Allahabad 211019, India}
\affiliation[b]{Regional Centre for Accelerator-based Particle Physics \\
                Harish-Chandra Research Institute, Allahabad 211019, India} 

\emailAdd{asesh@hri.res.in, nishiwaki@hri.res.in, sourabh@hri.res.in}
\preprint{HRI-P-12-06-001 \\ 
\vspace*{-0.8cm}
\begin{flushright}
RECAPP-HRI-2012-007
\end{flushright}
}
\abstract{We work out the strongly interacting sector of a non-minimal Universal 
Extra Dimension (nmUED) scenario with one flat extra spatial dimension orbifolded on 
$S^1/Z_2$ in the presence of brane-localized kinetic and Yukawa terms. 
On compactification, these terms are known to have significant, nontrivial 
impact on the masses and the couplings of the Kaluza-Klein (KK) excitations.
We study the masses of the level `1' KK gluon and the quarks and find the modified
strong interaction vertices involving these particles. The scenario conserves 
KK parity. Possibility of significant level-mixing among the quarks from different 
KK-levels is pointed out with particular reference to the top quark sector.
Cross sections for various generic final states involving level `1' KK-gluon and 
KK-quarks from first two generations are estimated at the Large Hadron Collider 
(LHC) via an implementation of the scenario in MadGraph-5 with the help of FeynRules. 
The decay branching fractions of both strong and weakly interacting KK excitations 
are studied to estimate yields in various different final states involving
jets, leptons and missing energy. 
{These are used to put some conservative constraints on the nmUED
parameter space using the latest LHC data.}
Nuances of the scenario are elucidated with reference to the minimal Universal Extra 
Dimension (mUED) and Supersymmetry (SUSY) and their implications {for} the LHC are 
discussed.} 

\begin{document} 
\maketitle

%\end{document}
%\kill
%
%%%%%%%%Section #1  %%%%%%%%%%%%%%%%%%%%%%%%%%%%%%%%%%%
%%%%%%%%%%%%%%%%%%%%%%%%%%%%%%%%%%%%%%%%%%%%%%%%%%%%%%%
\section{Introduction}
\label{sec:intro}

With the Large Hadron Collider (LHC) running to perfection for over two years
now, the high energy physics community is waiting for the first genuine
hint of coveted new physics with bated breath. While the nature of new physics that
would eventually be uncovered at the LHC can be anybody's guess, the amount of effort
that has gone in hypothesizing, modelling and analyzing the same over past
three decades is simply breathtaking. Out of these, supersymmetry (SUSY) and 
compactified extra spatial dimensions stand out as the two most generic and popular frameworks
for going beyond the Standard Model (SM) of particle physics. 
However, even within these two broad frameworks, these do not exhaust the possibilities. 
Thus, in an era when the hunt is on at the LHC for signatures of these popular scenarios,
studying in detail the newer possibilities assumes a special significance.

During the past decade, scenarios with TeV-scale extra dimensions \cite{Antoniadis:1990ew} 
has received serious attention. Scenarios with Universal Extra Dimensions (UED), 
first proposed in \cite{Appelquist:2000nn}, belong to this class where the 
extra-dimensional bulk has an \emph{universal} (indiscriminate) access to all the SM particles. 
The {simplest} version of such a scenario is a direct extension of the SM with only one extra, 
flat spatial dimension, orbifolded on
$S^1/Z_2$ and has two free parameters: the radius of compactification ($R$)
and the 
Higgs mass. The scenario,
with its chosen orbifold and appropriate orbifold-boundary conditions, ensures
the presence of chiral fermions of the SM and respects a $Z_2$ symmetry.
The latter, in turn, provides a stable dark matter candidate 
while ensuring that the Kaluza-Klein (KK) excitations of the orbifolded theory 
{that have} odd $Z_2$-parity (the KK-parity) always appear in pair at any interaction
vertex~\cite{Servant:2002aq, Cheng:2002ej, Kakizaki:2005en, Kakizaki:2005uy, Burnell:2005hm, Kong:2005hn, Matsumoto:2005uh, Kakizaki:2006dz, Matsumoto:2007dp, Belanger:2010yx}. 

However, such a scenario leads to an almost degenerate particle
spectrum at each KK level. Hence, this cannot have an interesting phenomenology,
particularly, at the colliders. On the other hand, being a 5-dimensional (5D) theory and hence 
being non-renormalizable, this can only be an effective theory characterized by a 
cut-off scale $\Lambda$. Higher order corrections to the KK masses are thus
inevitable and these involve $\Lambda$ \cite{Cheng:2002iz} which becomes the third
free parameter of the scenario. The loop-corrections lift the 
degeneracy of these masses thus opening the door for a rich phenomenology 
involving the KK particles. This is the scenario known in the literature as
the minimal UED (mUED).

Notwithstanding the importance of such a scenario be studied on its 
own merits, it has been shown to masquerade as SUSY in its signals at 
colliders \cite{Cheng:2002ab}. Thus, studying UED has acquired an extra facet.
Naturally, on top of generic collider studies of the scenario 
\cite{Macesanu:2002db, Carone:2003ms, Bhattacharyya:2005vm, Cembranos:2006gt, 
Bhattacherjee:2007wy, Bhattacherjee:2008ik, Konar:2009ae, Matsumoto:2009tb, 
Bhattacharyya:2009br, Bandyopadhyay:2009gd, Bhattacherjee:2010vm, Murayama:2011hj, 
Ghosh:2010tp, Nishiwaki:2011gm, Datta:2011vg, Ghosh:2012zc}
that also include constraining the parameter space of some of its variants from recent
LHC data \cite{Aad:2010qr, Nishiwaki:2011vi, Nishiwaki:2011gk, Belanger:2012zg, Kakuda:2012px}, 
discriminating SUSY from mUED has since been an active 
area of research \cite{Battaglia:2005zf, Datta:2005zs, Datta:2005vx}. The issue has triggered 
another important area of intense study, \emph{viz.}, measuring the spins
\cite{Battaglia:2005zf, Smillie:2005ar} 
of the new excitations at the (hadron) colliders which has been advocated 
to be the (only) direct way to resolve the confusion.

It is to be noted at this point that UED being an effective theory in 
4 space-time dimensions (4D), one needs to take into account all operators that
are allowed by the gauge symmetry of SM and Lorentz invariance. The possible
places where these can appear is the bulk or at the orbifold fixed-points,
\emph{i.e.}, the boundaries of the bulk and the brane.\footnote{
The presence of  these terms is not a mere question of a possibility.
It has been argued in the literature that they are genuinely required for the 
framework to be consistent.}
These are not \emph{a priori}
{un}known quantities (expected to be related to the {dynamics} of (yet to be understood) ultra-violet (UV) completion
for such scenarios) and thus would serve as extra free parameters of the 
theory.
In {the} case of mUED, all boundary-localized terms (BLTs) are assumed to vanish at the scale
$\Lambda$ only to be generated radiatively at the low scale that ultimately contributes
as corrections to the masses \cite{Cheng:2002iz}. It is important to note that in such a simplified 
extension of the SM, the interaction strengths of the level `1' KK particles remain 
unaltered when compared to the corresponding SM ones.

Set against such a backdrop, it is necessary to explore UED beyond a particular 
(and literally minimal) version with more generic set{-}ups that contain bulk
and boundary localized terms (BLTs). The mUED, with only two 
(three, considering the Higgs mass) free
parameters may still play the role of a very economic 
benchmark scenario like what minimal supergravity (mSUGRA) is to SUSY studies.
Fortunately, meantime, there have been important contributions discussing the 
theoretical possibilities
and {the plausible} structures with such terms. These can broadly be classified in two categories:
one where {the} bulk mass terms for the fermions are considered leading to the so-called
split-UED scenario \cite{Park:2009cs} whose phenomenology has been studied 
in some details in recent years \cite{Chen:2009gz, Kong:2010xk, Flacke:2011nb, Huang:2012kz} at colliders
and in connection to non-collider aspects like dark matter etc.
The other variety is the one in which non-vanishing boundary-localized kinetic 
terms at a high scale are considered 
\cite{Dvali:2001gm, Carena:2002me, delAguila:2003bh, delAguila:2003kd,
delAguila:2003gu, delAguila:2003gv, Muck:2004br, del Aguila:2006kj}. 
However, phenomenological
studies with BLTs have been scarce and perhaps, to the best of our knowledge,
ref.\cite{Flacke:2008ne} is the first {such} work that considered such terms. 
This work deals with the spectrum 
and couplings in the electroweak sector, their effects on ElectroWeak Symmetry Breaking (EWSB)  
and the interesting possibility of having multiple dark matter candidates. 
Recently, LHC data have been used to constrain such an electroweak sector \cite{Flacke:2012ke}.
Note that both split-UED and the scenarios with BLTs mentioned above have in-built
mechanisms that {help} preserve KK-parity and thus could provide suitable dark matter
candidate(s).
Also, an LHC-study \cite{Datta:2012xy} (perhaps, the first of its kind) of a 
UED scenario with unequal BLTs at the two orbifold fixed-points 
(that give rise to KK-parity violation) has recently been circulated.

In the present work, we take up a scenario with one flat, universal, extra spatial 
dimension. We concentrate on the effects of the boundary localized
kinetic terms (BLKTs) and the Yukawa terms (BLYTs) in the strongly 
interacting sector comprising of KK gluon and KK quarks and their basic 
phenomenology at the LHC. We limit ourselves only to the first KK level 
except for the top quark sector for which we briefly discuss some interesting 
aspects involving the second KK level. We also restrict our analysis to 
a common BLT term at the two orbifold fixed points for both gluon and the quarks 
thus preserving the KK-parity. The BLT terms for these two sectors are expressed
in terms of two (three, including BLYT {which is} important for the top quark sector) 
mutually independent parameters that serve as the only two (three)
additional ones {when} compared to the mUED case. In order to carry out a parton-level 
study of the final states with jets+leptons+$\slashed E_T$ we incorporate
an electroweak sector which is mUED-like.

Further, in this work, we do not consider the 
effect of radiative corrections to the KK masses thus leaving out $\Lambda$
as a parameter.
As we would see later in this paper, BLTs can indeed generate much
larger splitting among gluon and quarks at the first KK level than what radiative 
corrections could inflict in mUED, for a given value of $\rinv$. 
In that sense and in the spirit of ref.\cite{Flacke:2008ne}, this analysis has a complementary
aspect to that in mUED. Hence, the scenario we work in has three relevant parameters:
$\rinv$, $r_G$ and $r_Q$, the latter two being the BLT parameters for the KK gluon and 
the quarks respectively (along with the mass of the Higgs boson). 

We observe that BLTs can indeed inflict major distortions in the mUED spectrum beyond
recognition \cite{Datta:2005zs, Datta:2005vx}. On top of that, some of the crucial 
couplings involving the KK quarks, gluons and the electroweak gauge bosons are
modified in a nontrivial way. This can not only alter the (mUED) expectations 
at the LHC in a characteristic way, but also could open up new possibilities.
Thus, such a framework would provide a rather relaxed framework which can make 
the confusion among mUED, nmUED,
and SUSY (and also possibly, $T$-parity conserving little Higgs framework 
(see refs.~\cite{Schmaltz:2005ky, Perelstein:2005ka} and references therein){)}
get more complete.

The paper is organized as follows.
In section \ref{sec:framework} we discuss the theoretical framework that include 
the BLTs for the strongly interacting sector indicating 
their nontrivial implications. In section \ref{sec:spectrum} we derive the mass 
spectrum and the couplings and highlight their features by contrasting them with 
those in the mUED framework. The resulting phenomenology at the LHC is taken up 
in section \ref{sec:lhc} where we discuss in detail the production rates 
of the level `1' KK gluon and quarks and the decay branching fractions
of the KK particles involved in the cascades as functions of the 
fundamental parameters of the framework. Situations in nmUED are studied with 
concrete examples to demonstrate the possibility of a near-complete faking from 
mUED and SUSY scenarios. Some characteristic discriminators that could partially
alleviate the confusion under favourable conditions are also discussed with
reference to various different final states at the LHC.
In section \ref{sec:outlook} we conclude.
We also provide an appendix for the Feynman rules involving the interactions
of the strongly interacting KK particles from level `1'
{which are used in this work.}

\section{Theoretical framework}
\label{sec:framework}
We consider the strongly interacting (QCD) sector of a 5D UED scenario compactified
on $S^1/Z_2$ in the presence of brane-localized terms. Under a $Z_2$ orbifold on 
$S^1$, two fixed points appear and some 4D terms, consistent with gauge symmetry and 
Lorentz invariance, can be localized around them. Theoretical aspects of 
brane-localized kinetic terms (BLKTs) have been studied {in great details}
refs.~\cite{Dvali:2001gm, Carena:2002me, delAguila:2003bh, delAguila:2003gu, 
delAguila:2003gv, Muck:2004br, del Aguila:2006kj}.
We follow the notations of ref.~\cite{Flacke:2008ne}, where a UED scenario with 
brane localized terms only for the electroweak gauge bosons and Higgs sectors 
(and not for the gluon and the fermion sectors) are considered. 
The total action for the QCD sector can be expressed as:
\beq
S_{\text{NMQCD}} = S_{\text{quark}}  + S_{\text{gluon}} + S_{\text{Yukawa}},
\label{SNMQCD}
\eeq
where the different components of the complete action are as follows:
\al{              
S_{\text{quark}} &= \int d^4 x \int_{-L}^{L} dy \sum_{i=1}^3 \Bigg\{
                i \overline{U}_i \Gamma^{M} \mathcal{D}_{M} U_i + r_Q \Big( \delta(y-L) + \delta(y+L) \Big)
                \Big[ i \overline{U}_i \gamma^{\mu} \mathcal{D}_{\mu} P_L U_i \Big] \notag \\
&\phantom{= \int d^4 x \int_{-L}^{L} dy \sum_{i=1}^3 \!}
                +i \overline{D}_i \Gamma^{M} \mathcal{D}_{M} D_i + r_Q \Big( \delta(y-L) + \delta(y+L) \Big)
                \Big[ i \overline{D}_i \gamma^{\mu} \mathcal{D}_{\mu} P_L D_i \Big] \notag \\
&\phantom{= \int d^4 x \int_{-L}^{L} dy \sum_{i=1}^3 \!}
                +i \overline{u}_i \Gamma^{M} \mathcal{D}_{M} u_i + r_Q \Big( \delta(y-L) + \delta(y+L) \Big)
                \Big[ i \overline{u}_i \gamma^{\mu} \mathcal{D}_{\mu} P_R u_i \Big] \notag \label{Squark} \\
&\phantom{= \int d^4 x \int_{-L}^{L} dy \sum_{i=1}^3 \!}
                +i \overline{d}_i \Gamma^{M} \mathcal{D}_{M} d_i + r_Q \Big( \delta(y-L) + \delta(y+L) \Big)
                \Big[ i \overline{d}_i \gamma^{\mu} \mathcal{D}_{\mu} P_R d_i \Big] \Bigg\}, \\
S_{\text{gluon}} &= \int d^4 x \int_{-L}^{L} dy \Bigg\{ - \frac{1}{4} G^{a}_{MN} G^{aMN} +
                \Big(\delta(y-L) + \delta(y+L) \Big) \Big[ - \frac{r_G}{4} G^{a}_{\mu\nu} G^{a\mu\nu} \Big] \Bigg\},
                \label{Sgluon}
}
\al{         
S_{\text{Yukawa}} &= \int d^4 x \int_{-L}^{L} dy \sum_{i,j=1}^3 \Bigg\{ -\Big( 1+ r_Y \paren{\delta(y-L) + \delta(y+L)} \Big) \notag \\
& \hspace{40mm}              \times \Big[ Y^{u}_{ij} \overline{Q}_i u_j \tilde{\Phi} +Y^{d}_{ij} \overline{Q}_i d_{j} \Phi +\text{h.c.}
                \Big] \Bigg\}. \label{SYukawa}
}
In the above set of expressions, $y$ represents the compact extra spatial direction;
$M, N$ run over $0,1,2,3,y$ while $\mu, \nu$ run over $0,1,2,3$. Representations of 
the 5D Minkowski metric and the Clifford algebra are chosen as 
$\eta_{MN} = \text{diag}(1,-1,-1,-1,-1)$ and $\Gamma^M = \{ \gamma^{\mu}, i\gamma^5 \}$, respectively.
The 4D chiral projectors for the {right and the left-handed states have the usual definition of} 
$P_{R,L} = \frac{1 \pm \gamma^5}{2}$.
$G^{a}_{M}, U_i, D_i, u_i, d_i, \Phi$ correspond to {the} 5D gluon, {the} 5D up- and down-type 
$SU(2)_W$ doublet quarks from the $i$-th generation, the same for the $SU(2)_W$ 
singlet quarks, $SU(2)_W$ Higgs doublet, respectively. {`$a$'} is the $SU(3)_C$ adjoint index.
$Y^{u}_{ij}$ and $Y^{d}_{ij}$ are the 5D Yukawa {matrices}. 
$\tilde{\Phi}$ respects the condition $\tilde{\Phi} = i\sigma_2 \Phi^{\ast}$ 
with $\sigma_2$ being the conventional Pauli matrix.
Concrete forms of the {5D} covariant derivative for the fermions ({$\mathcal{D}_M$}) and
{the field strength} for the gluon field are given by
\al{
\mathcal{D}_M &= \pal_M -i g_{_{5s}} G^a_{M} T^a, \\
G^{a}_{MN} &= \pal_M G^a_{N} - \pal_N G^a_{M} + g_{_{5s}} f^{abc} G^{b}_{M} G^{c}_{N},
}
where $g_{_{5s}}$ is the 5D strong (QCD) coupling, $T^a$ is {the} $SU(3)_C$ generator from 
the fundamental representation and $f^{abc}$ is the $SU(3)_C$ group structure constant.

In this paper, we consider the {so-called} {``downstairs" } picture where we only 
focus on the fundamental region of the $Z_2$-orbifold extended over $[-L, L]$ with 
$L=\pi R/2$, $R$ being the radius of the compact extra dimension~\cite{Haba:2009uu}. The $Z_2$ orbifolding 
leads to a discrete symmetry in the extra-dimensional ($y$) coordinate that can be
expressed as 
\al{
{y+L \sim - (y+L)}
}
with two fixed points at $y=\pm L$.
{The 5D covariant forms of the brane-localized terms in equations (\ref{Sgluon}),\,(\ref{Squark}),\,(\ref{SYukawa}) 
can be shown to have their 4D counterparts which do not break the gauge symmetry of {the QCD sector}}. 
We assume that the electroweak gauge symmetry is spontaneously broken by the ordinary Higgs 
mechanism as it is in the case of mUED. It is noted that the Vacuum Expectation Value (VEV) of the
Higgs field can possess a constant profile (even in the presence of brane-localized Higgs terms that have
covariant forms in 4D) by tuning appropriate parameters~\cite{Flacke:2008ne}.\footnote{Another possibility of theories with non-constant ($y$-dependent) Higgs VEV have been pursued in refs.~\cite{Coradeschi:2007gb,Burgess:2008ka,Haba:2009uu,Fujimoto:2011kf}.}
Note that the total action in equation (\ref{SNMQCD}) is invariant under the transformation 
$y \rightarrow -y$ which exchanges the positions of the two fixed points.
This suggests that the theory has an accidental $Z_2$ symmetry, called KK-parity, 
which ensures the stability of the lightest KK particle thus making the same a viable dark matter candidate.
All these issues are taken up in the subsequent sections in reference to 
KK excitations of the gluon and the quarks. This we do by discussing first the 
`free' parts in the respective actions and thereby determining the profiles of the
corresponding KK excitations. Subsequently, we focus on the `interaction' parts of the
actions, presented earlier, involving the gluon(s) and the quarks.

%%%%%%%%%%%%%%%%%%%
\subsection{Free parts of gluon and quark}
%%%%%%%%%%%%%%%%%%%

{The forms of the bulk equations of motion and the boundary conditions at the 
two orbifold fixed points (located at $y=\pm L$) are determined using variational principle
following ref.~\cite{Csaki:2003dt,Csaki:2003sh}.
In this paper we use the unitary gauge with $G^a_y \rightarrow 0$ where $G^a_y$
are unphysical degrees of freedom.}
Here, $\Psi$ ($\psi$) represents the `up' and `down'-type KK quark fields of the {$SU(2)_W$} 
doublet (singlet), $U_i, D_i$ ($u_i, d_i$) where {`$i$'}
stands for flavour. However, we do not distinguish between quark flavours 
because the structure of $S_{\text{quark}}$ in equation (\ref{Squark}) is flavour-blind.

In the unitary gauge, the 5D fields of $G^a_{\mu}, \Psi, \psi$ are KK-decomposed as follows:
\al{
G^a_{\mu}(x,y) &= \sum_{n=0}^{\infty} G^{a(n)}_{\mu}(x) f_{G_{(n)}}(y),
\label{5Dgluondecomposition} \\
\Psi_L(x,y) &= \Psi^{(0)}_L(x) f_{\Psi_{(0)L}}(y) + \sum_{n>0:\text{even}}
                \Psi^{(n)}_L(x) f_{\Psi_{(n)L}}(y) + \sum_{n>0:\text{odd}}
                \Psi^{(n)}_L(x) f_{\Psi_{(n)L}}(y), \\
\Psi_R(x,y) &= \sum_{n>0:\text{even}}
                \Psi^{(n)}_R(x) f_{\Psi_{(n)R}}(y) + \sum_{n>0:\text{odd}}
                \Psi^{(n)}_R(x) f_{\Psi_{(n)R}}(y), \\
\psi_R(x,y) &=    \psi^{(0)}_R(x) f_{\psi_{(0)R}}(y) + \sum_{n>0:\text{even} }
                \psi^{(n)}_R(x) f_{\psi_{(n)R}}(y) + \sum_{n>0:\text{odd} }
                \psi^{(n)}_R(x) f_{\psi_{(n)R}}(y),   \\
\psi_L(x,y) &=    \sum_{n>0:\text{even} }
                \psi^{(n)}_L(x) f_{\psi_{(n)L}}(y) + \sum_{n>0:\text{odd} }
                \psi^{(n)}_L(x) f_{\psi_{(n)L}}(y),
}
where the mode functions of level-$n$ states can be categorized as
\al{
f_{G_{(n)}}(y) &= N_{G_{(n)}} \times
\begin{cases}
\displaystyle \frac{\cos(M_{G_{(n)}} y)}{C_{G_{(n)}}}  & \text{for $n$ even} \\
\displaystyle \frac{{-}\sin(M_{G_{(n)}} y)}{S_{G_{(n)}}}  & \text{for $n$ odd}
\label{gluonmodefunction}
\end{cases}, \\
f_{Q_{(n)}} \; \equiv \; f_{\Psi_{i(n)L}} = f_{\psi_{i(n)R}}  &= N_{Q_{(n)}} \times
                \begin{cases}
                \displaystyle \frac{\cos(M_{Q_{(n)}} y)}{C_{Q_{(n)}}}  & \text{for $n$ even} \\
                \displaystyle \frac{{-}\sin(M_{Q_{(n)}} y)}{S_{Q_{(n)}}}  & \text{for $n$ odd}
                \end{cases},  \label{eqn:f-profiles} \\
g_{Q_{(n)}} \; \equiv \; f_{\Psi_{i(n)R}} = -f_{\psi_{i(n)L}} &= N_{Q_{(n)}} \times
                \begin{cases}
                \displaystyle \frac{\sin(M_{Q_{(n)}} y)}{C_{Q_{(n)}}}  & \text{for $n$ even} \\
                \displaystyle \frac{\cos(M_{Q_{(n)}} y)}{S_{Q_{(n)}}}  & \text{for $n$ odd} 
                \end{cases}, \label{eqn:g-profiles}
}
with the normalization factors $N_{G_{(n)}}, N_{Q_{(n)}}$.
Hereafter, we use the following short-hand notations
\al{
C_{X_{(n)}} =\cos\paren{\frac{M_{X_{(n)}}\pi R}{2}},\quad
S_{X_{(n)}} =\sin\paren{\frac{M_{X_{(n)}}\pi R}{2}},\quad
T_{X_{(n)}} =\tan\paren{\frac{M_{X_{(n)}}\pi R}{2}},
\label{trigabbrevi}
}
where $X$ stands for $G$ (gluon) and $Q$ (quark), and $M_{X_{(n)}}$ is the corresponding 
KK mass {at the $n$-th level determined through the transcendental equations}
\al{
r_X {M_{X_{(n)}}} =
\begin{cases}
- T_{X_{(n)}} & \text{for $n$ even} \\
1/T_{X_{(n)}} & \text{for $n$ odd}
\label{gluonKKmasscondition}
\end{cases}.
}
The generalized orthonormal conditions for $\{ f_{X_{(n)}} \}$ and $\{ g_{Q_{(n)}} \}$ take the forms
\begin{equation}
\begin{array}{rc}
\displaystyle \int_{-L}^{L} dy \Big[ 1 + r_X \paren{\delta(y-L) + \delta(y+L)} \Big] f_{X_{(m)}} f_{X_{(n)}} &= \delta_{m,n}, \\
\displaystyle \int_{-L}^{L} dy\,  g_{Q_{(m)}} g_{Q_{(n)}} &= \delta_{m,n},
\end{array}
\label{Psiorthonormalrelation}
\end{equation}
respectively, while the expressions for {$N_{X_{(n)}}$} {turn out to be as follows:}
\al{
N_{X_{(n)}}^{-2} =
\begin{cases}
\displaystyle 2r_X + \frac{1}{C_{X_{(n)}}^2} \left[ \frac{\pi R}{2} + \frac{1}{2M_{X_{(n)}}} \sin(M_{X_{(n)}} \pi R) \right] & \text{for $n$ even} \\
\displaystyle 2r_X + \frac{1}{S_{X_{(n)}}^2} \left[ \frac{\pi R}{2} - \frac{1}{2M_{X_{(n)}}} \sin(M_{X_{(n)}} \pi R) \right] & \text{for $n$ odd}
\end{cases}{.}
\label{Psinormalization}
}
Note that, in the presence of BLTs, these normalization-factors have rather nontrivial forms when compared to the simple
forms like $1 \over \sqrt{\pi R}$ or $1 \over \sqrt{2 \pi R}$ as in the case of mUED.
Especially, the profile for the zero mode is normalized as
\al{
N_{X_{(0)}} = \frac{1}{\sqrt{2r_X + \pi R}},
\label{eqn:NG0}
}
which results in the following theoretical lower bound on $r_X$ in order to circumvent a tachyonic zero mode: 
\al{
r_X > - \frac{\pi R}{2}.
}

%%%%%%%%%%%%%%%%%%%%%%%%%%%%%%
\subsection{The Yukawa sector and the quark mass matrix}
\label{sec:yukawa}
%%%%%%%%%%%%%%%%%%%%%%%%%%%%%%

{The Yukawa sector of such an nmUED scenario has previously been considered in
ref. \cite{delAguila:2003kd} where its implications for electroweak precision data were discussed.
In this work, we work out the salient features of this sector with particular
reference to the masses and the couplings of the KK quarks from the third generation.}

On EWSB via the ordinary Higgs mechanism, the Higgs doublet $\Phi$ acquires the VEV $\langle \Phi \rangle = (0, v/\sqrt{2})^{\text{T}}$ with $v = 246\,\text{GeV}$.
We assume that the brane-localized Yukawa terms are flavour-blind thereby allowing us to diagonalize 
the Yukawa matrices $Y^{u}_{ij}$ and $Y^{d}_{ij}$ in a way similar to that {for} the SM and which can {be} expressed as
\al{
\int d^4 x \int_{-L}^{L} dy \Bigg\{ -\Big( 1+ r_Y \paren{\delta(y-L) + \delta(y+L)} \Big)
                \sum_{i=1}^{3}\Big[ \Big( \mathcal{Y}^{u}_{ii} \frac{v}{\sqrt{2}} \Big) \overline{U}_i u_i
                + \Big( \mathcal{Y}^{d}_{ii} \frac{v}{\sqrt{2}} \Big) \overline{D}_i d_{i} +\text{h.c.}
                \Big] \Bigg\},
}
where $\mathcal{Y}^{u}_{ii}, \mathcal{Y}^{d}_{ii}$ are {the} diagonalized Yukawa matrices
and we concentrate on the mass terms.
Hereafter, we restrict ourselves to the first KK mode (unless otherwise {indicated})
and focus on the generic flavour `$i$' with $q_i$ {$(Q_i)$} representing the corresponding $SU(2)_W$ 
singlet {(doublet)}, respectively.
After some calculations, we obtain
\al{
- \Big( \mathcal{Y}^{q}_{ii} \frac{v}{\sqrt{2}} \Big) \int d^4 x \Bigg\{
R_{Q00} \overline{q}^{(0)}_{iL} q^{(0)}_{iR} + r_{Q11} \overline{Q}^{(1)}_{iL} q_{iR}^{(1)}
- R_{Q11}  \overline{q}^{(1)}_{iL} Q_{iR}^{(1)} + \text{h.c.} \Bigg\},
\label{Yukawaterms}
}
where{, for clarity,} we make a redefinition of $u^{(0)}_{iL} = U^{(0)}_{iL}${.} $R_{Q00}, r_{Q11}, R_{Q11}$ results {from} the overlap integral {and are given by}
\al{
R_{Q00} &= \int_{-L}^{L} dy \Big( 1+ r_Y \paren{\delta(y-L) + \delta(y+L)} \Big)
                f_{Q_{(0)}}^2 = \frac{2 r_Y + \pi R}{2 r_Q + \pi R},  \\
r_{Q11} &= \int_{-L}^{L} dy \Big( 1+ r_Y \paren{\delta(y-L) + \delta(y+L)} \Big)
                f_{Q_{(1)}}^2 \notag \\
&=  \frac{2r_Y + \frac{1}{S_{Q_{(1)}}^2} \left[ \frac{\pi R}{2} - \frac{1}{2M_{Q_{(1)}}} \sin(M_{Q_{(1)}} \pi R) \right] }
                {2r_Q + \frac{1}{S_{Q_{(1)}}^2} \left[ \frac{\pi R}{2} - \frac{1}{2M_{Q_{(1)}}} \sin(M_{Q_{(1)}} \pi R) \right]}, \\
R_{Q11} &= \int_{-L}^{L} dy \Big( 1+ r_Y \paren{\delta(y-L) + \delta(y+L)} \Big)
                g_{Q_{(1)}}^2 \notag  \\
&= \frac{2r_Y ( C_{Q_{(1)}}/S_{Q_{(1)}} )^2 + \frac{1}{S_{Q_{(1)}}^2} \left[ \frac{\pi R}{2} + \frac{1}{2M_{Q_{(1)}}} \sin(M_{Q_{(1)}} \pi R) \right]}
{\frac{1}{S_{Q_{(1)}}^2} \left[ \frac{\pi R}{2} + \frac{1}{2M_{Q_{(1)}}} \sin(M_{Q_{(1)}} \pi R) \right]}.
}
The zero mode masses {(\emph{i.e.}, the masses of the SM quarks)} are fixed as
\al{
m_{q_i} = \Big( \mathcal{Y}^{q}_{ii} \frac{v}{\sqrt{2}} \Big) R_{Q00}.
}
It is noted that when $r_Y = - \pi R/2$, the value of $R_{Q00}$ becomes zero and the SM quarks
become massless. Obviously, this limit is meaningless in phenomenology and we should avoid {the} possibility.
On the other hand, in the limit $r_Q = r_Y$, values of both $R_{Q00}$ and $r_{Q11}$ become 1 {while} $R_{Q11}$ 
is still away from $1$. This implies that deviations from the mUED case 
may still be present in the physical mass spectrum of the level `1' KK quarks.
The mUED limit is recovered with $r_G = r_Q =0$ when all of $R_{Q00}, r_{Q11}, R_{Q11}$ become equal to $1$.
This, in turn, implies that non-vanishing $r_Y$ may play some role in determining even the spectrum of the KK
quarks that correspond to the lighter flavours of the SM. The effect is generally miniscule 
{for their mass-eigenvalues} 
since equation (\ref{Yukawaterms}) has an overall factor {which amounts to}
the mass of the SM quark of $i$-th light flavour. Exception to this for the top
quark sector will be pointed out at the end of section \ref{sec:masses}.
{However, as we will find later, the Yukawa sector has an important implication
for the mixing between the weak eigenstates of the KK quarks of lighter SM flavours.}

{There is another interesting phenomenon known as \emph{level-mixing} that can take place
between similar states from two different KK levels{.} 
This explicitly violates KK number. However, this is perfectly admissible since {the}
translational invariance in 5D {is broken at the orbifold fixed points} which is otherwise synonymous with the idea of
KK number conservation. 
However, to conserve KK-parity, {the mixings} would be limited
to {those} between two even or two odd states only. 
The possibility of such \emph{level-mixings} has already been pointed out in the literature but
its phenomenological implications are yet to be {explored} thoroughly in various different contexts. 
{In the case of mUED, such effects can only be induced at a higher order.} However, presence
of BLTs ensures overlap integrals of the following form:
\al{
\int_{-L}^{L} dy \Big( 1+ r_Y \paren{\delta(y-L) + \delta(y+L)} \Big) f_{Q_{(m)}} g_{Q_{(n)}},
\label{eqn:ry-overlap-integral}}
which triggers \emph{level-mixings} even at the tree-level for cases with
$(m, n)=$ (even, even) or (odd, odd).
Note that such effects are only possible when ${{Y}^{q}_{ij}} \ne 0${.}
The contribution is found to be negligible {though} for the first two generations but is not 
always so for the KK top quarks. We will {indicate} the phenomenological implications
of such mixing effects in the later part of this work. However, we postpone a somewhat
elaborate discussion on the issue to a future work. In any case,
for convenience at a later stage, we rewrite equation (\ref{Yukawaterms}) as follows:
\al{
- m_{q_i} \int d^4 x \Bigg\{
 \overline{q}^{(0)}_{iL} q^{(0)}_{iR} + r'_{Q11} \overline{Q}^{(1)}_{iL} q_{iR}^{(1)}
- R'_{Q11} \overline{q}^{(1)}_{iL} Q_{iR}^{(1)} + \text{h.c.} \Bigg\},
\label{Yukawaterms2}
}
where $r'_{Q11}, R'_{Q11}$ are defined as}
\al{
r'_{Q11} = \frac{r_{Q11}}{R_{Q00}}, \quad
R'_{Q11} = \frac{R_{Q11}}{R_{Q00}}.
}

Now we can obtain the mass matrix for the level `1' KK quarks as
\al{
- \int d^4 x \Bigg\{
\begin{bmatrix} \overline{Q}^{(1)}_i, \ \overline{q}^{(1)}_i \end{bmatrix}_L
\underbrace{
\begin{bmatrix}
                M_{Q_{(1)}} & r'_{Q11} m_{q_i} \\
                -R'_{Q11} m_{q_i} & M_{Q_{(1)}}
\end{bmatrix}}_{\equiv \mathcal{M}^{(1)}_{q_i}}
\begin{bmatrix} Q^{(1)}_i \\ q^{(1)}_i \end{bmatrix}_R
+ \text{h.c.} \Bigg\}.
\label{1stKKupmasses}
}
By choosing
same mass for these entries we implicitly assume that the BLKTs for the quarks are blind to $SU(2)_W$
quantum numbers (singlet or doublet) they possess.
By use of the following bi-unitary transformation of
\al{
\begin{bmatrix} Q^{(1)}_i \\ q^{(1)}_i \end{bmatrix}_L
= V^{(1)}_{q_iL}
\begin{bmatrix} \mathcal{Q}^{(1)}_{i2} \\ \mathcal{Q}^{(1)}_{i1} \end{bmatrix}_L, \quad
\begin{bmatrix} Q^{(1)}_i \\ q^{(1)}_i \end{bmatrix}_R
= V^{(1)}_{q_iR}
\begin{bmatrix} \mathcal{Q}^{(1)}_{i2} \\ \mathcal{Q}^{(1)}_{i1} \end{bmatrix}_R,
\label{biunitaryform_for_LandR}
}
we can diagonalize equation (\ref{1stKKupmasses}) as follows:
\al{
- \int d^4 x
\begin{bmatrix} \overline{\mathcal{Q}}^{(1)}_{i2}, \ \overline{\mathcal{Q}}^{(1)}_{i1} \end{bmatrix}
\begin{bmatrix}
                m^{(1)}_{q_i2} &  \\
                & m^{(1)}_{q_i1}
\end{bmatrix}
\begin{bmatrix} \mathcal{Q}^{(1)}_{i2} \\ \mathcal{Q}^{(1)}_{i1} \end{bmatrix},
}
where $\mathcal{Q}^{(1)}_{i1}, \mathcal{Q}^{(1)}_{i2}$ are the mass eigenstates of level `1' KK quarks.
The set of eigenvalues, {$\paren{m^{(1)}_{q_i1}}^2, \paren{m^{(1)}_{q_i2}}^2$}, of the mass matrix squared
$\mathcal{M}^{(1)}_{q_i} \mathcal{M}^{(1)\dagger}_{q_i}$ are assumed with $m^{(1)}_{q_i2} > m^{(1)}_{q_i1}$.
The forms of the matrices $V^{(1)}_{q_iL}, V^{(1)}_{q_iR}$ are fixed by the eigenvectors of 
$\mathcal{M}^{(1)}_{q_i} \mathcal{M}^{(1)\dagger}_{q_i}$ simultaneously.
{The profiles of level 1 top and bottom quark masses are illustrated in section \ref{sec:spectrum}}.

%%%%%%%%%%%%%%%%%%%%%%%%%%%%%%%%%%%%%%%%%%%%%%%%%%%%%
%%%%%%%%%%%%%%%%%%%%%%%%%%%%%%%%%%%%%%%%%%%%%%%%%%%%%
%\section{Properties of masses and interactions under parameters in branes}
\section{Mass spectrum and couplings}
\label{sec:spectrum}
%%%%%%%%%%%%%%%%%%%%%%%%%%%%%%%%%%%%%%%%%%%%%%%%%%%%%
%%%%%%%%%%%%%%%%%%%%%%%%%%%%%%%%%%%%%%%%%%%%%%%%%%%%% 

In this section we discuss the variations of the masses of the level `1' 
KK quarks and the KK gluon and the dependence of the strength of the
interaction between them as a function of $\rinv$ and parameters 
like $r_G$, $r_Q$ and $r_Y$.
For convenience, the latter three dimensionful parameters are rescaled
in terms of $R$ as shown below.
\al{
r'_G = r_G \rinv, \quad
r'_Q = r_Q \rinv, \quad
r'_Y = r_Y \rinv,
}
It is to be noted that, with this redefinition, the {variables}
$C, S$ and $T$ in equations (\ref{trigabbrevi}) now become functions of {scaled mass parameters}
{$M'_{G_{(n)}}$} ($M'_{Q_{(n)}}$) instead of {$M_{G_{(n)}}$} ($M_{Q_{(n)}}$),
respectively. We define {and use} these modified mass parameters in the 
subsections to follow. 

%%%%%%%%%%%%%%%%%%%%%%%%%%%%
\subsection{Masses of level `1' KK gluon and quarks}
\label{sec:masses}
%%%%%%%%%%%%%%%%%%%%%%%%%%%%

\begin{figure}[h]
\begin{center}
\includegraphics[width=0.49\columnwidth]{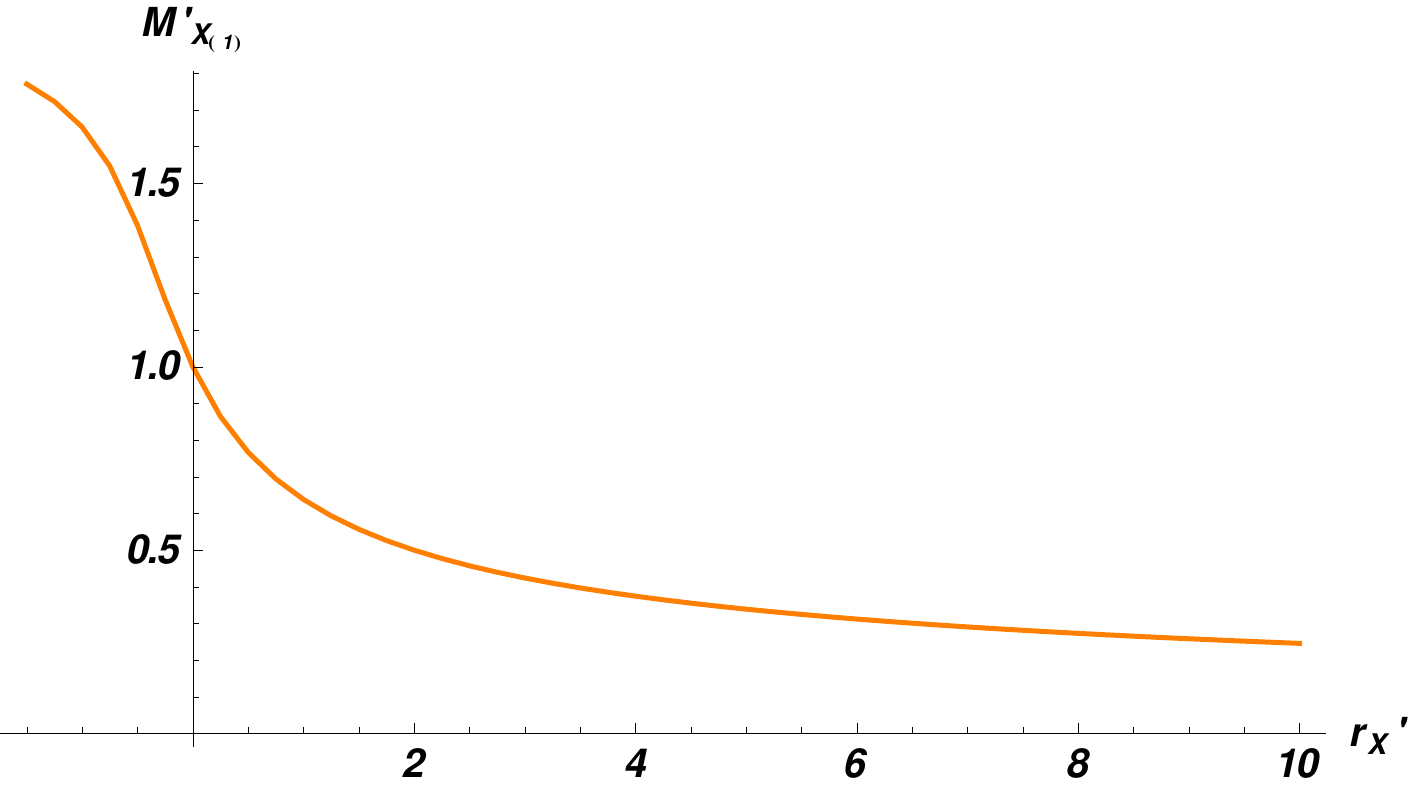}
%\hspace*{1.0cm}
\includegraphics[width=0.49\columnwidth]{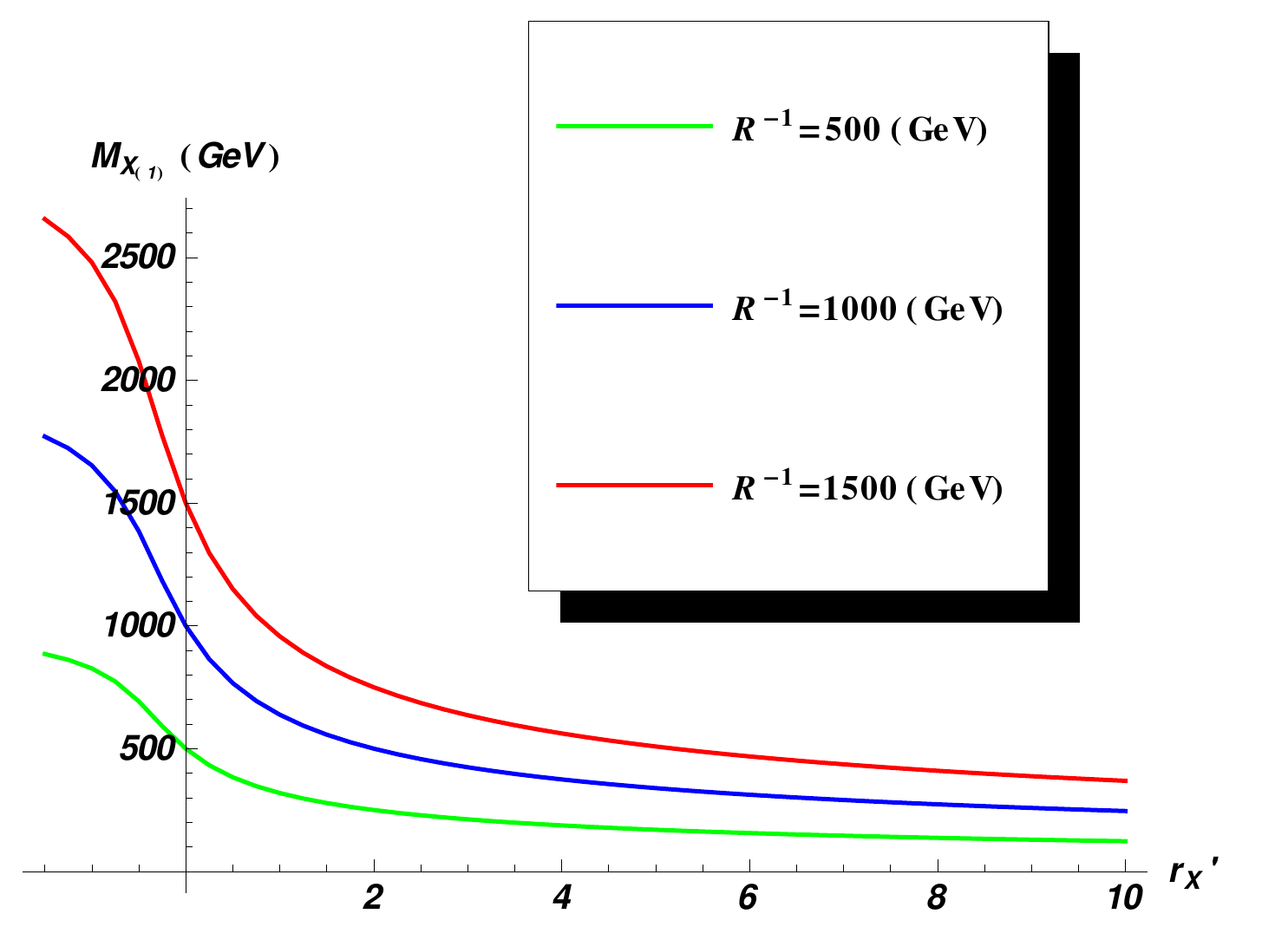}
\end{center}
\caption{Ratio of actual {KK} mass of level `1' KK gluon/quark and $\rinv$
(left panel) and the corresponding actual masses (right panel; for different 
values of $\rinv$) plotted against the parameter {$r'_X$ characterizing} the brane-localized 
term. The trivial case of {$M'_{X_{(1)}}=1$} (left panel) or 
{$M_{X_{(1)}}=\rinv$} (right panel) is retrieved when $r'_X=0$.}
\label{KKmassdeviationratio_pdf}
\end{figure}

From equations (\ref{gluonKKmasscondition})
one finds that KK masses of both level `1' gluon and quarks (from the
first two generations) are governed by identical set of transcendental
equations involving $r'_G$ (for the KK gluon) and $r'_Q$ (for the KK quarks).
 However, this statement is true only at the lowest
order. Radiative corrections to the masses would be different for the
KK quarks and the KK gluons but estimating the same is beyond the scope 
of the present work.
{The transcendental equations for the odd `$n$' (for level `1' KK-gluon and quark) from expressions} 
(\ref{gluonKKmasscondition}) and 
can be rewritten in terms of the scaled 
variables $r'_G$ ($r'_Q$) and {$M'_{G_{(1)}}$} ($M'_{Q_{(1)}}$) as follows
\al{
r'_{X} {M'_{X_{(1)}}} = 1/T_{X_{(1)}} \label{masscondition},
}
where {$M'_{X_{(1)}} = M_{X_{(1)}}/R^{-1}$} and $X$ stands for $G$ {(gluon)} and $Q$ {(quark)}.
{These} transcendental equations are solved numerically for the {KK} masses of
the level `1' KK gluon and quarks. The variations of the masses are plotted 
in {figure}~\ref{KKmassdeviationratio_pdf} as a function of $r'_X$.

 By virtue of 
equation (\ref{masscondition}),
this dependence is blind to $\rinv$. It is interesting to note 
that for $r'_{X} < 0$, {$M'_{X_{(1)}} > 1$} signifying the actual KK mass to be
larger than $\rinv$. The reverse is true for $r'_{X} > 0$. {As we can see}
from this panel that the variation flattens up quickly with increasing 
$r'_{X}$.

\begin{figure}[h]
\centering
\includegraphics[width=75mm, clip]{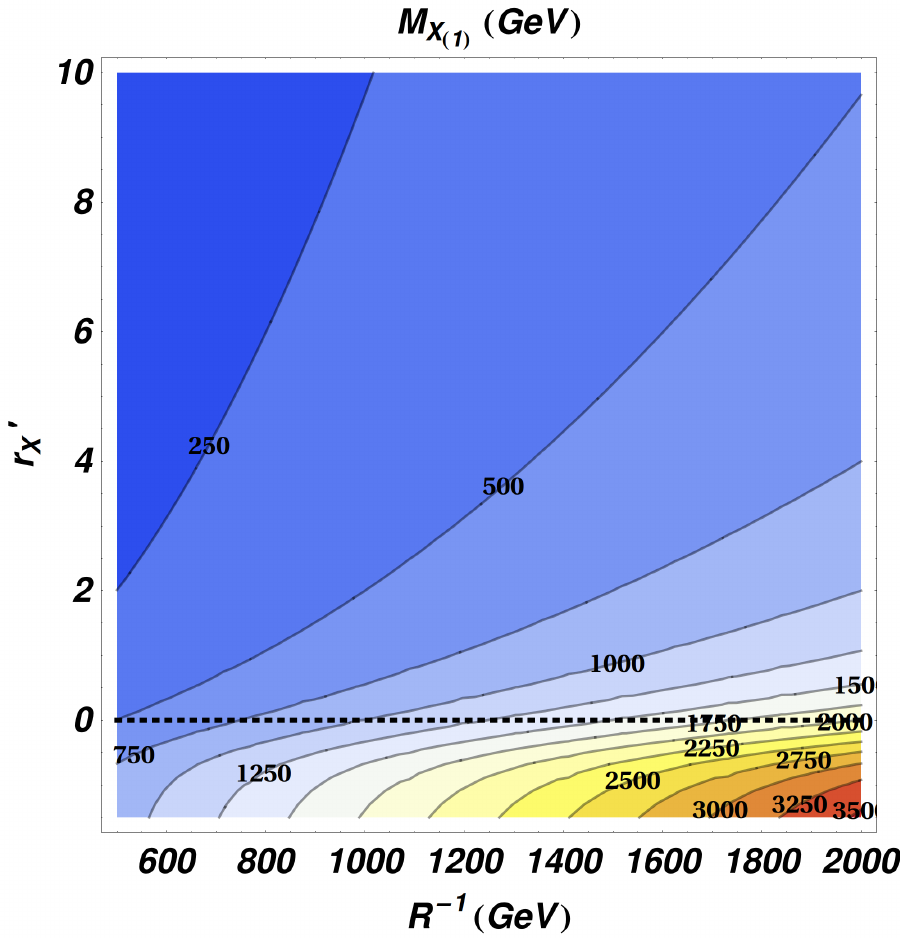}
\caption{{\emph{IsoKKmass}} (in GeV) contours for level `1' KK gluon/quark in the
$\rinv-r'_X$ plane.}
\label{isomass}
\end{figure}

In the right panel of figure \ref{KKmassdeviationratio_pdf} we show the
actual variations of {KK} masses (\emph{i.e.}, of {$M_{X_{(1)}}$}) for level `1' KK gluon/quark 
for three given values of $\rinv$. This plot readily follows from the one 
in the left panel using the relation between {$M_{X_{(1)}}$} and {$M'_{X_{(1)}}$}
as indicated above. This also reveals that a particular mass-value for
the KK gluon (quark) could arise from different combinations of
$\rinv$ and $r'_G$ ($r'_Q$) which is further illustrated in figure
\ref{isomass} for a continuous range of $\rinv$.

{This leads us to explore the \emph{isomass} contours in the $\rinv-r'_X$ plane 
as illustrated in figure \ref{isomass}.} 
This shows clearly how similar values of {KK} masses can be obtained for different combinations
of $\rinv$ and $r'_X$. Note that the straight line represented by $r'_G, r'_Q=0$
(parallel to {the} $\rinv$-axis) cuts the mass contours at values of $\rinv$ equal
to the mass-value of the contour. This is in conformity with 
figure \ref{KKmassdeviationratio_pdf}. 

We give a quantitative summary for the {KK} masses of level `1' KK gluon/quark
in table~\ref{somevaluesofKKgluon} by providing some concrete numbers.
{$M'_{X_{(1)}}$} represents the solutions of equation (\ref{masscondition})
for reference input values of $r'_X$ which are independent of $\rinv$
(as discussed earlier in this subsection). The actual {KK} masses are simple
products of {$M'_{X_{(1)}}$} and $\rinv$. One such set of actual masses is 
shown for $\rinv=1000$ GeV in table~\ref{somevaluesofKKgluon}.

In the above discussion we have taken a simplistic approach as far as the 
masses of the level `1' KK quarks are concerned. It should be kept in mind
that the mass-eigenvalues of the KK quarks would {be evaluated from 
$\mathcal{M}^{(1)}_{q_i} \mathcal{M}^{(1)\dagger}_{q_i}$ in equation
(\ref{1stKKupmasses})}. In general, the two eigenvalues are not degenerate
because of the presence of non-vanishing overlap integrals like $r'_{Q11}$,
$R'_{Q11}$ {\it etc.} which are by themselves dimensionless and are also governed
by dimensionless parameters like $r'_Q$, $r'_Y$, $M'_{Q_{(1)}}$ {\it etc.} 
When contrasted with mUED, this is 
a clear new feature appearing in the framework of UED with brane-localized 
terms. However, as can be seen from equation {(\ref{1stKKupmasses})}, the
mass-splitting is proportional to the value of the corresponding zero-mode
quark mass and thus negligible for the level `1' KK quarks from the first two 
generations. In this limit, the mass eigenvalues ($m^{(1)}_{{q_i}_{(1,2)}}$)
becomes identical to the KK mass ({$M_{Q_{(1)}}$}).
{On the contrary, $M_{G_{(1)}}$ corresponds to the physical mass of $G^{(1)}$.}

%\vskip 20pt

\begin{table}[H]
\begin{center}
\begin{tabular}{|c|c||c|}
\hline
$r'_X$ & {$M'_{X_{(1)}}$} & ${M_{_{X_{(1)}}}} \,\text{(GeV)}$ \\
       &              &   (for $R^{-1}=1000\,\text{GeV}$) \\         
\hline
-1.5 & 1.771 & 1771 \\
-1.0 & 1.654 & 1654 \\
-0.5 & 1.386 & 1386 \\
0.0 & 1.000 & 1000 \\
0.5 & 0.767 & 767 \\
1.0 & 0.638 & 638 \\
2.0& 0.500 & 500 \\
5.0 & 0.339 & 339 \\
10.0 & 0.246 & 246 \\
\hline
\end{tabular}
\caption{{KK masses} for level `1' KK gluon/quark{s} for varying $r'_X$ and for $\rinv=1000$ GeV.}
\label{somevaluesofKKgluon}
\end{center}
\end{table}

\begin{figure}[h]
\centering
\includegraphics[width=0.49\columnwidth, clip]{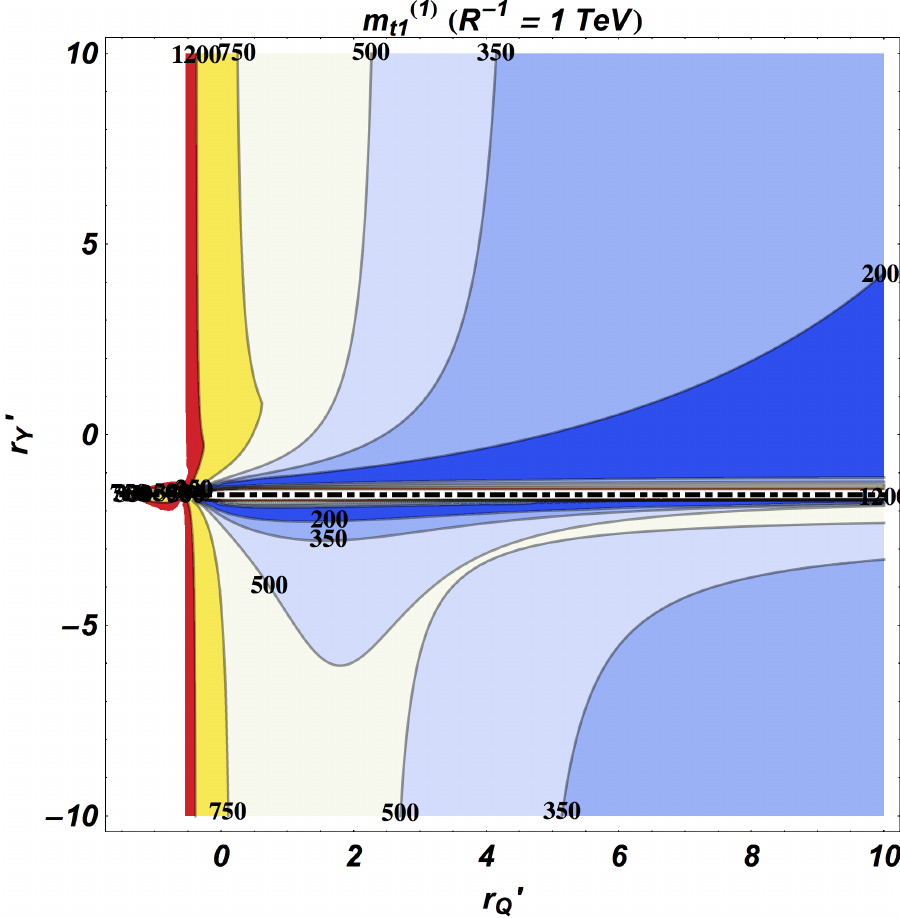}
%\hspace*{1.0cm}
\includegraphics[width=0.49\columnwidth, clip]{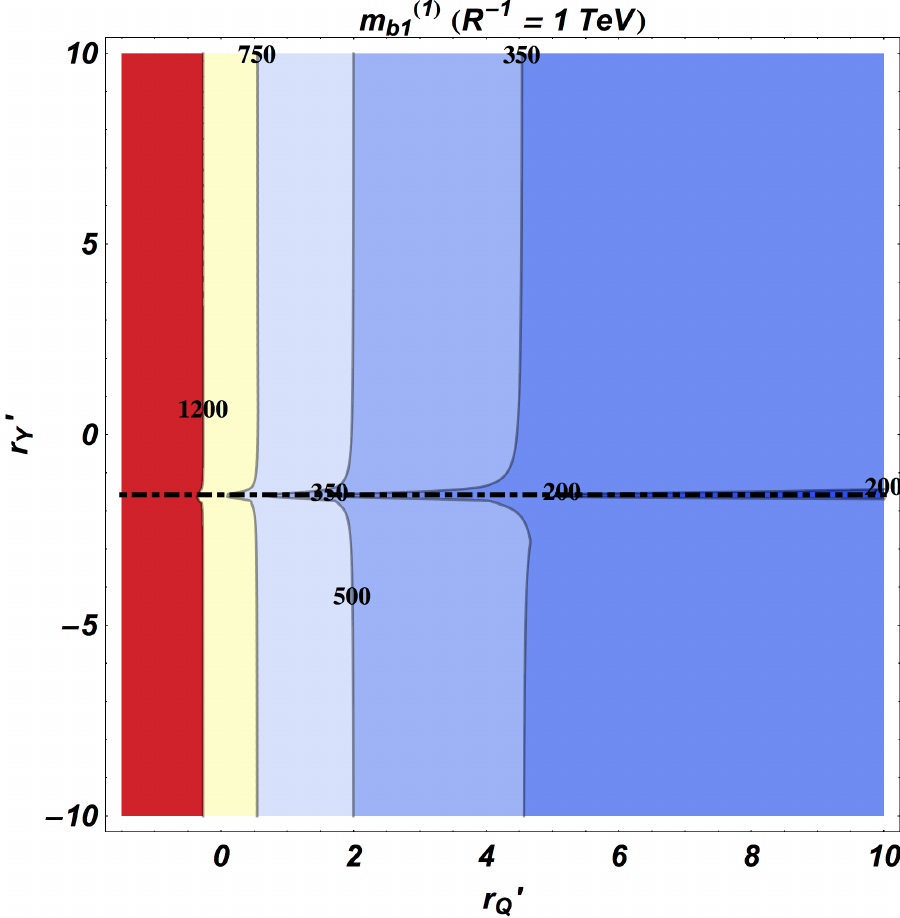}
\caption{Isomass contours for the light level `1' top quark (left) and the light 
level `1' bottom quark (right) 
for $R^{-1} = 1000\,\text{GeV}$ in the $r'_Q-r'_Y$ plane.} 
\label{fig:KK1-top-bot-1000_pdf}
\end{figure}

The phenomenon is not quite unexpected though since the effect under 
consideration originates in the Yukawa sector of the theory. Thus, such an
effect will be appreciable for only the KK top quarks and to a far lesser extent
for the {KK} bottom quarks. In the left plot of figure
\ref{fig:KK1-top-bot-1000_pdf} we illustrate the effect for the lighter top quark {with the help of \emph{isomass} 
contours} that show significant, nontrivial
dependence of the mass on $r'_Y$ in addition to that on $r'_Q$ for a given value 
of $\rinv$ (= 1000 GeV). 
The right panel of figure \ref{fig:KK1-top-bot-1000_pdf} is for the case of
lighter level `1' KK bottom quark. This one clearly reveals that for level `1'
KK quarks corresponding to the lighter SM quarks, the dependence of their masses
on $r'_Y$ is small. Thus, these two plots collectively help one estimate the quantitative 
role of $r'_Y$ in the phenomenon. 

As discussed in the beginning of {section \ref{sec:yukawa}}, at around 
$r'_Y= -\pi/2$ the values of $r'_{Q11}, R'_{Q11}$ rise sharply and get divergent. 
In both plots of figure \ref{fig:KK1-top-bot-1000_pdf},
this results in a thin strip of region about this value of $r'_Y$  over 
which there is no {physical} solution. Further, as mentioned at the end of section 
\ref{sec:yukawa}, because of the extremal situation it can lead to, 
$r'_Y$ close to its limiting value of $-\pi/2$ can have non-trivial bearing 
even on the properties of the KK quarks of light flavours, at least,
in principle.

Further, the possibility of level-mixing between similar KK-parity states
driven by the brane-localized Yukawa term (as noted from equation 
(\ref{eqn:ry-overlap-integral})) emerges as an interesting feature of the top 
quark sector whose phenomenology could be rather rich in such a scenario. 
Our preliminary investigation into the subject reveals that mixing
between level `0' and level `2' top quarks can be \emph{a priori} significant.
Such a mixing could potentially trigger
an appreciable shift in the mass of the level `2' top quark and make the same phenomenologically
interesting at the LHC. Moreover, as could be expected, the SM top mass receives
contribution from such a mixing. Thus, refined experimental estimates of the
mass of the SM top quark from Tevatron~\cite{Brandt:2012ui} and the LHC~\cite{Silva:2012di} would inevitably constrain
the parameters of the nmUED scenario we are considering here. A detailed 
study of the sector involving the KK quarks from the third generation including
the role of level-mixing is beyond the scope of this paper and would be taken
up separately in a future work.

%%%%%%%%%%%%%%%%%%%%%%%%%%%%
\subsection{Interactions involving level `1' KK gluons and quarks}
\label{sec:couplings}
%%%%%%%%%%%%%%%%%%%%%%%%%%%%
%
In this subsection we discuss the other important aspect of the framework,
\emph{viz.}, the couplings involving the KK gluon and KK quarks. Here {again},
we limit ourselves only to the first KK level.

 4D QCD coupling $g_{4s}$ is defined as
\al{
g_{_{4s}} \equiv N_{G_{(0)}} g_{_{5s}} = \frac{g_{_{5s}}}{\sqrt{2r_G + \pi R}}{.}
\label{4DQCDcouplingdefinition}
}
Quartic interaction involving four level `1' KK gluons is somewhat non-trivial and gets modified by the 
presence of brane-localized terms. {However, it is rather inconsequential for LHC phenomenology and hence
we do not discuss this any further.} 
All other self-coupling terms involving level `1' KK gluon and SM gluon
(both 3-point and 4-point ones) remain the same as in mUED.

Next, we turn to the case of the interaction involving a level `1' 
KK gluon and a level `1' KK quark along with an {(level `0')} SM quark. 
Here we comment on the forms of {the} bi-unitary matrices 
$V^{(1)}_{q_iL}$ and $V^{(1)}_{q_iR}$  that diagonalize the mass 
matrix for level-1 KK quarks where {`$i$' refers to 
the quark-flavour}.
%%%%% Start Here
For (almost) mass-degenerate KK quarks (in the limit of $r'_Q=r'_Y$ 
which we adopt for studying the KK quarks corresponding to
lighter SM flavours), $V^{(1)}_{q_iL}$ and $V^{(1)}_{q_iR}$ can be shown, to a very
good approximation, to have the following form {that} reflects maximal mixing: 
\al{
V^{(1)}_{q_iL} = V^{(1)}_{q_iR} \; \approx
                \begin{bmatrix}
                - \text{sgn}(r'_Q) \cos\paren{\frac{\pi}{4}} & \sin\paren{\frac{\pi}{4}} \\
                - \text{sgn}(r'_Q) \sin\paren{\frac{\pi}{4}} & - \cos\paren{\frac{\pi}{4}}
                \end{bmatrix},
\label{eqn:Vmatrix1}
}
except for the case of $r'_Q=0$.\footnote{This general form of the matrix is used in our 
subsequent analysis. It should be noted that this expression is qualitatively different 
from its mUED counterpart for which it is an unit matrix {(see equation~(\ref{mUEDlimitingbiunitarymatrix}))} and this cannot be seen as a limiting
case (\emph{i.e.}, $r'_G=r'_Q=0$) of the former.}
In the case of conventional UED scenarios without brane-localized terms, 
one finds the mass-eigenvalues to be exactly degenerate ({before} radiative correction to the masses)
and these matrices look like:
\al{
 V^{(1)}_{q_iL}|_{\text{mUED}}  =
                \begin{bmatrix}
                \cos(\theta^{(1)}_{q_i}) & \sin(\theta^{(1)}_{q_i}) \\
                -\sin(\theta^{(1)}_{q_i}) & \cos(\theta^{(1)}_{q_i})
                \end{bmatrix}, \quad
 V^{(1)}_{q_iR}|_{\text{mUED}}  =
                \begin{bmatrix}
                \cos(\theta^{(1)}_{q_i}) & -\sin(\theta^{(1)}_{q_i}) \\
                \sin(\theta^{(1)}_{q_i}) & \cos(\theta^{(1)}_{q_i})
                \end{bmatrix},
\label{eqn:Vmatrix-mued}
}
which include chiral rotation and the mixing angle $\theta^{(1)}_{q_i}$ is fixed by
\al{
\sin(2 \theta^{(1)}_{q_i}) = \frac{m_{q_i}}{\sqrt{M_{Q_{(1)}}^2 + m_{q_i}^2}}, \quad
\cos(2 \theta^{(1)}_{q_i}) = \frac{M_{Q_{(1)}}}{\sqrt{M_{Q_{(1)}}^2 + m_{q_i}^2}}{,}
\label{eqn:sinth-costh}
}
{where $m_{q_i}$ is the mass of the `$i$' th flavour SM quark.}
{The} difference in form of the matrices presented in
equations (\ref{eqn:Vmatrix1}) and (\ref{eqn:Vmatrix-mued}) owes its origin to
the difference between `approximate degeneracy' and `exact degeneracy' of the mass-eigenvalues
of the quarks.
Further, it may be noted that for the five light flavours, $M_{Q_{(1)}} \gg m_{q_i}$.
Thus, use of equation (\ref{eqn:sinth-costh}) reduces equation (\ref{eqn:Vmatrix-mued})
to the following trivial form:
\al{
V^{(1)}_{q_iL}|_{\text{mUED}}  = V^{(1)}_{q_iR}|_{\text{mUED}}  =
		\begin{bmatrix}
		1 & 0 \\
		0 & 1
		\end{bmatrix}
		\quad ({\text{\emph{i.e.,}}} \; \theta^{(1)}_{q_i} = 0),
		\label{mUEDlimitingbiunitarymatrix}
}
whose form is different from that of equation~(\ref{eqn:Vmatrix1}).

Using equation (\ref{Squark}), the 4D effective action depicting the quark-gluon
interaction up to the first KK level can be written down as follows:
\al{
S_{\text{quark}} |_{\text{int}} &=
                \int d^4 x \sum_{{i}} \Bigg\{ g_{4s} T^{a} \Bigg[ G_{\mu}^{a(0)} \Big(
                \overline{q}^{(0)}_{i} \gamma^{\mu} q_{i}^{(0)} + \overline{\mathcal{Q}}^{(1)}_{i1} \gamma^{\mu} \mathcal{Q}_{i1}^{(1)} + \overline{\mathcal{Q}}^{(1)}_{i2} \gamma^{\mu}\mathcal{Q}_{i2}^{(1)} \Big) 
\notag \\
&\hspace{-17mm} 
+G_{\mu}^{a(1)} (g'_{G_1Q_1Q_0}) \Bigg(
                \overline{q}^{(0)}_{i} \gamma^{\mu} \paren{ v^{(1)}_{q_i R(21)} P_R + v^{(1)}_{q_i L(11)} P_L } \mathcal{Q}_{i2}^{(1)} + \overline{q}^{(0)}_{i} \gamma^{\mu} \paren{ v^{(1)}_{q_i R(22)} P_R + v^{(1)}_{q_i L(12)} P_L } \mathcal{Q}_{i1}^{(1)} 
\notag \\
&   
+\overline{\mathcal{Q}}^{(1)}_{i2} \gamma^{\mu} \paren{ v^{(1)}_{q_i R(21)} P_R + v^{(1)}_{q_i L(11)} P_L } q_{i}^{(0)} + \overline{\mathcal{Q}}^{(1)}_{i1} \gamma^{\mu} \paren{ v^{(1)}_{q_i R(22)} P_R + v^{(1)}_{q_i L(12)} P_L } q_{i}^{(0)} \Bigg) \Bigg] \Bigg\}  
\label{eqn:s-quark-int},
}
where the superscripts $0,1$ in parenthesis indicate the KK level{.}
$\mathcal{Q}_{1,2}$ represent the quark mass-eigenstates at the first KK level, 
$i$ is the {generic} flavour-index and $v_q$-s 
are the elements of the $V_q$ matrices in equations (\ref{biunitaryform_for_LandR}),\,(\ref{eqn:Vmatrix1}). The latter can
now be rewritten in the following general form:
\al{
V^{(1)}_{q_iL} =
                \begin{bmatrix}
                v^{(1)}_{q_i L(11)} & v^{(1)}_{q_i L(12)} \\
                v^{(1)}_{q_i L(21)} & v^{(1)}_{q_i L(22)}
                \end{bmatrix}, \quad
V^{(1)}_{q_iR} =
                \begin{bmatrix}
                v^{(1)}_{q_i R(11)} & v^{(1)}_{q_i R(12)} \\
                v^{(1)}_{q_i R(21)} & v^{(1)}_{q_i R(22)}  \label{eqn:Vmatrix2}
                \end{bmatrix}.
}
{The first term in equation (\ref{eqn:s-quark-int}) gives the usual interaction
of the SM gluon with a pair of SM quarks.
The next two terms give the 
interactions of the SM gluon with two different pairs of mass-eigenstates of level `1' 
KK quarks and these are identical to their mUED counterparts.
This is because they are governed by the overlap integral
\al{
\int_{-L}^{L} dy \Big( 1+ r_Q \paren{\delta(y-L) + \delta(y+L)} \Big) f_{G_{(0)}} f_{Q_{(1)}} f_{Q_{(1)}},
}
{which reduces to $f_{G_{(0)}}(= N_{G_{(0)}}$, the normalization factor 
in equation~(\ref{4DQCDcouplingdefinition})) by virtue of the manifest identity 
in equation~(\ref{Psiorthonormalrelation}).}
The only deviation that occurs
is in the case of an SM quark interacting with a level `1' KK quark and a level `1' KK gluon.
The concrete form of the deviation (with respect to the mUED case) can be shown
to be as in equation (\ref{deviation-factor}).
\al{
g'_{G_1Q_1Q_0} &   \equiv \frac{1}{N_{G(0)}} \int_{-L}^{L} dy \Big( 1+ r_Q \paren{\delta(y-L) + \delta(y+L)} \Big) f_{G_{(1)}} f_{Q_{(1)}} f_{Q_{(0)}}   \notag \\
&\hspace{-15mm} =
  \frac{N_{Q_{(0)}}}{N_{G_{(0)}}} \frac{N_{G_{(1)}} N_{Q_{(1)}}}{S_{G_{(1)}} S_{Q_{(1)}}} \Big[
2r_Q S_{G_{(1)}} S_{Q_{(1)}} - \frac{\sin( (M_{Q_{(1)}} + {M_{G_{(1)}}}) \frac{\pi R}{2})}{M_{Q_{(1)}} + {M_{G_{(1)}}}} + \frac{\sin( (M_{Q_{(1)}} - {M_{G_{(1)}}}) \frac{\pi R}{2})}{M_{Q_{(1)}} - {M_{G_{(1)}}}} \Big]. 
                \label{deviation-factor}
}

The factor $g'_{G_1Q_1Q_0}$ is dimensionless and hence does not depend upon $\rinv$ which is a dimensionful parameter. In fact,
$g'_{G_1Q_1Q_0}$ is implicitly governed by the 
dimensionless parameters $r'_G$ and $r'_Q$ through the variables appearing in 
equation (\ref{deviation-factor}). This is a rather complicated dependence
and its concrete profile has a rich structure as shown in figure~\ref{coup-rg-rq}.
In the limit $r_G = r_Q$, it can be shown that $g'_{G_1Q_1Q_0}=1$ which is the mUED.

\begin{figure}[H]
\centering
\includegraphics[width=65mm, height=65mm , clip]{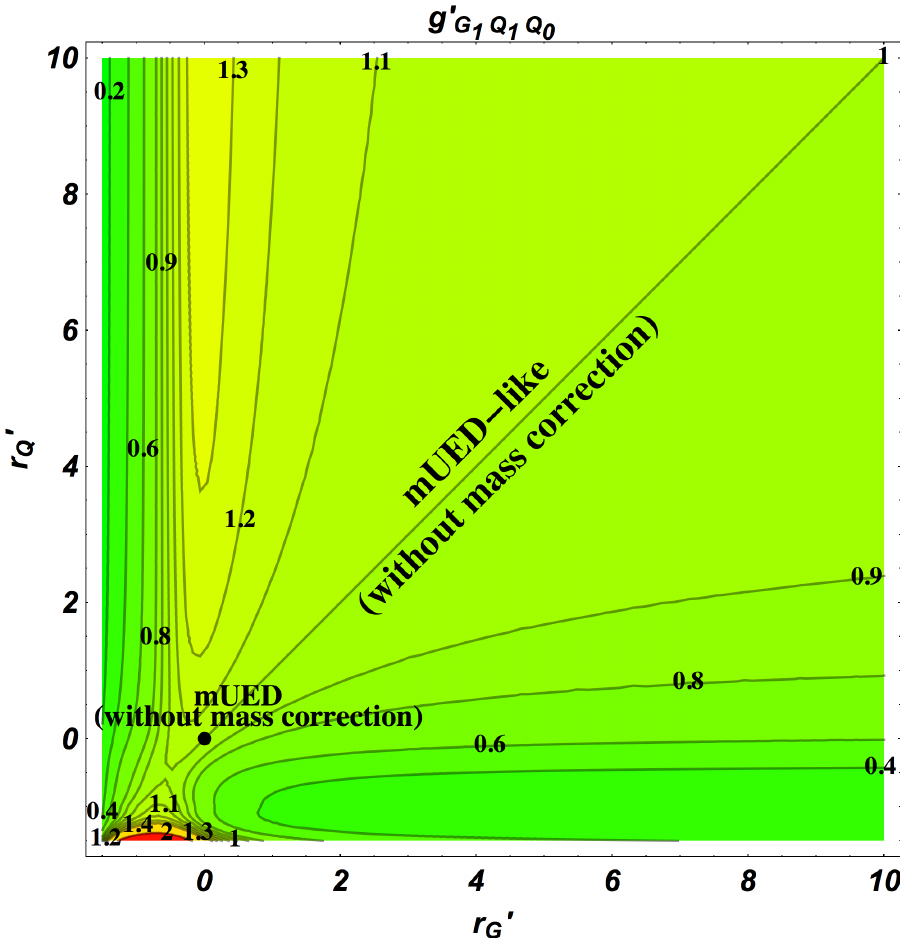}
\hspace*{1cm}
\includegraphics[width=65mm, height=65mm , clip]{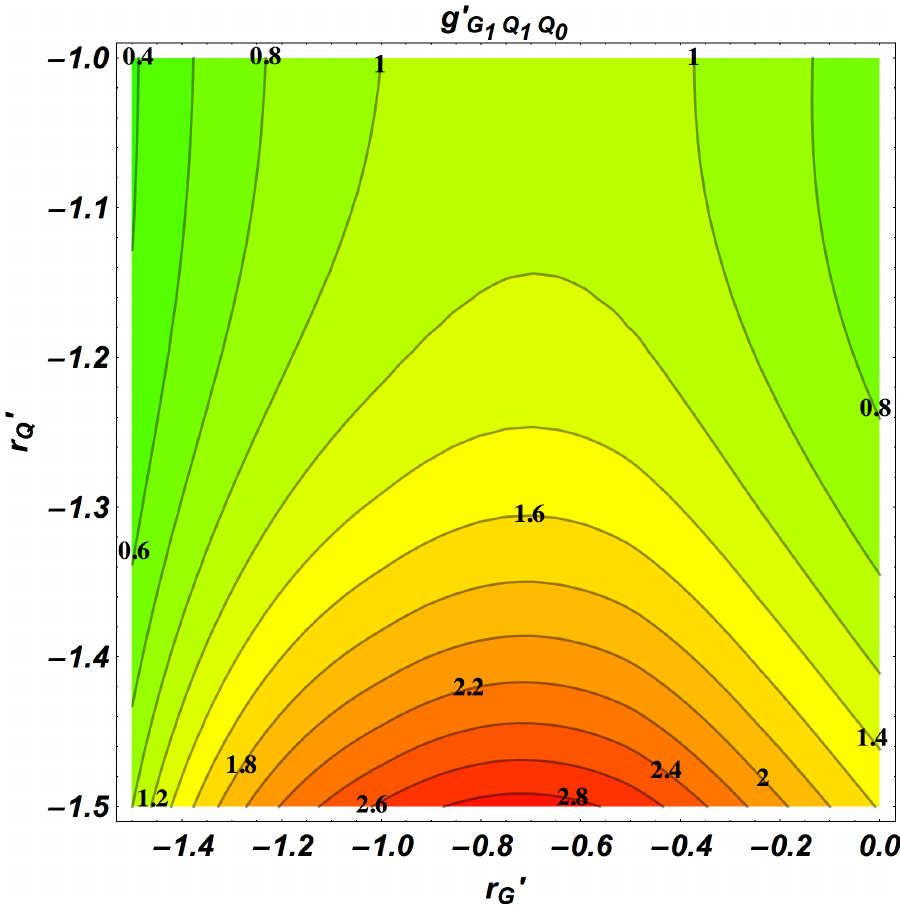}
\caption{Contours of deviation in $G_1 Q_1 Q_0$ coupling in nmUED with respect
to the mUED case: over larger ranges of values for $r'_G$ and $r'_Q$ (left)
and a zoomed up view over ranges of negative values for both (right) with
interesting variations. Note that ${r'_{G_{min}}}={r'_{Q_{min}}} = -1.5$ 
for these plots. This is somewhat above the theoretical minimum of $-{\pi \over 2}$ 
for both the parameters for which the scenario
{becomes} unphysical (see text for details).} 
\label{coup-rg-rq}
\end{figure}

In figure \ref{coup-rg-rq} we present the contours of the deviation factor
$g'_{G_1Q_1Q_0}$ presented in equation (\ref{deviation-factor}) in the $r'_G-r'_Q$
plane. The figure on left illustrates the contours over larger ranges of values 
for $r'_G$ and $r'_Q$. It is to be noted that along its diagonal ($r'_G=r'_Q$)
the deviation is exactly equal to 1 implying the coupling to be equal to that
in the mUED.\footnote{{The scenarios residing on the diagonal thus have 
degenerate KK masses which are different from those expected in a UED scenario 
without BLT (loosely indicated as mUED in the plot) for any given value of 
$R^{-1}$.}
We already assumed that in general, BLTs contribute dominantly to the KK masses 
when compared to the radiative corrections. The mUED scenario is defined only 
with the latter ones. Hence, on the diagonal, the scenarios are ``mUED-like''.} 
{The coupling has a much richer structure at very low values of 
$r'_G$ and $r'_Q$ close to the origin of the figure (indicated by blots
in red and yellow) as
both parameters approach their limiting value of $-\frac{\pi}{2} (\simeq -1.56)$.}
This is perhaps best 
understood if we just look at the form $\frac{N_{Q_{(0)}}}{N_{G_{(0)}}}$ in equation
(\ref{deviation-factor}) for which
both $N_{Q_{(0)}}$ and $N_{G_{(0)}}$ blow up at the said value.
{A closer look into this region is offered by a zoomed-up view in 
the right frame  of figure \ref{coup-rg-rq}. }

To probe further into the generic aspects of correlated variations of the
KK masses and the deviations in coupling from the mUED value, it would be
useful to follow up with a study showing their mutual variation. This is
pertinent since, as indicated above, the masses of the KK-quark and KK-gluon 
(we restrict ourselves to level `1' KK excitations only) are also functions
of $r'_G$ and $r'_Q$ as does the deviation-factor. The only difference is
that while the masses do vary {with} $\rinv$, the deviation-factor does not.

Thus, analogous to figure \ref{coup-rg-rq}, contours of fixed deviations in 
the couplings can be drawn but this time in the {$M_{G_{(1)}}-M_{Q_{(1)}}$} plane 
with $\rinv$ as a parameter. Such variations are shown in figure 
\ref{coup-mg-mq}. In the top panel of figure \ref{coup-mg-mq}, from left to
right, we present the case of $\rinv=1$ TeV and 2 TeV while in the bottom 
panel the corresponding ones illustrate the cases for $\rinv=3$ TeV and 5 TeV,
respectively. In order to facilitate the correspondences between the brane
parameters and the masses of the respective excitations for different values
of $\rinv$, the ranges of $r'_G$ (along the abscissa) and $r'_Q$ 
(along the ordinate) are indicated on the top and the right of each of these plots.
In both cases, the diagonal represents the contour for
$g'_{G_1Q_1Q_0}=1$. Under the hood, the geometrical origin of the diagonal 
has a common thread to that in the left panel of figure \ref{coup-rg-rq},
\emph{i.e.}, for $r'_G=r'_Q$, although the ranges considered for them are
different from the earlier case. 
The small region in yellow and red close to the top-right corner of the 
{top-left plot in figure \ref{coup-mg-mq}} corresponds 
to the bottom-left corner of the left plot in figure \ref{coup-rg-rq}.

For figure \ref{coup-mg-mq}, the criteria for choosing the mass-ranges for the 
level `1' gluon and quarks are, in turn, primarily based on the tentative reach 
of LHC ($\sim 3$ TeV) running at the centre of mass energy of 14 TeV and then, 
choosing not too large values of $r'_G$ and $r'_Q$ for different values 
of $\rinv$ considered for these plots. Recall that, in the scenario we are 
considering in the present work, equal values of {$M_{G_{(1)}}$ 
and $M_{Q_{(1)}}$}, for a given $\rinv$ result from equal values of $r'_G$ and 
$r'_Q$, respectively. 
Thus, as is clearly seen from figure \ref{coup-mg-mq}, 
degenerate masses occur along the diagonal.
As pointed out in the context of figure \ref{coup-rg-rq}, here also, by the 
same token, mUED-like scenarios live close along the diagonals.

Figure \ref{coup-mg-mq} tells us that different combinations of masses for
level `1' gluon and level `1' quarks would correspond to very specific values
of the deviation-factor for the modified coupling. The deviation can go 
either way, \emph{i.e.}, $g'_{G_1 Q_1 Q_0} \gtrless 1$. However, the 
correspondence  between masses and the deviation in coupling is specific to 
the value of $\rinv$, as can be understood by comparing the plots presented in figure~\ref{coup-mg-mq}. 
We like to emphasize that this correspondence, in principle, could be exploited
at the LHC to extract information on the parameters of the scenario. 
For example, if the masses in context can be known and the relevant cross 
sections can be estimated from the data, these could be used
to determine the deviation in coupling.\footnote{Extracting a somewhat precise 
information about the deviation in coupling could be a challenging task at 
a hadron collider. This is because any attempt to understand this from a 
total yield (where all production processes contribute) would inevitably 
involve the decay-patterns 
of the originally produced new-physics excitations. Extracting some concrete
information from within such a milieu requires further assumptions over the
scenario and thus, the exercise {may become} heavily `model-dependent'. However, the situation is expected
to be much under control in an extremely constrained scenario like the mUED
where the production cross sections could very well be related to the decay-patterns
of the produced particles. This is the case with us since we are trying to
measure a deviation of the coupling from its mUED value.} 
Since this deviation, when combined,
with the information on the masses, has a unique relationship to $\rinv$
in the current scenario, the latter can also be determined {subsequently}. The information
thus obtained on $\rinv$, in turn, can be employed to determine the values
of $r'_G$ and $r'_Q$ since these determine the masses which are{, by now,} known.

\vskip -10pt
\begin{figure}[H]
\centering
\includegraphics[width=0.49\columnwidth , clip]{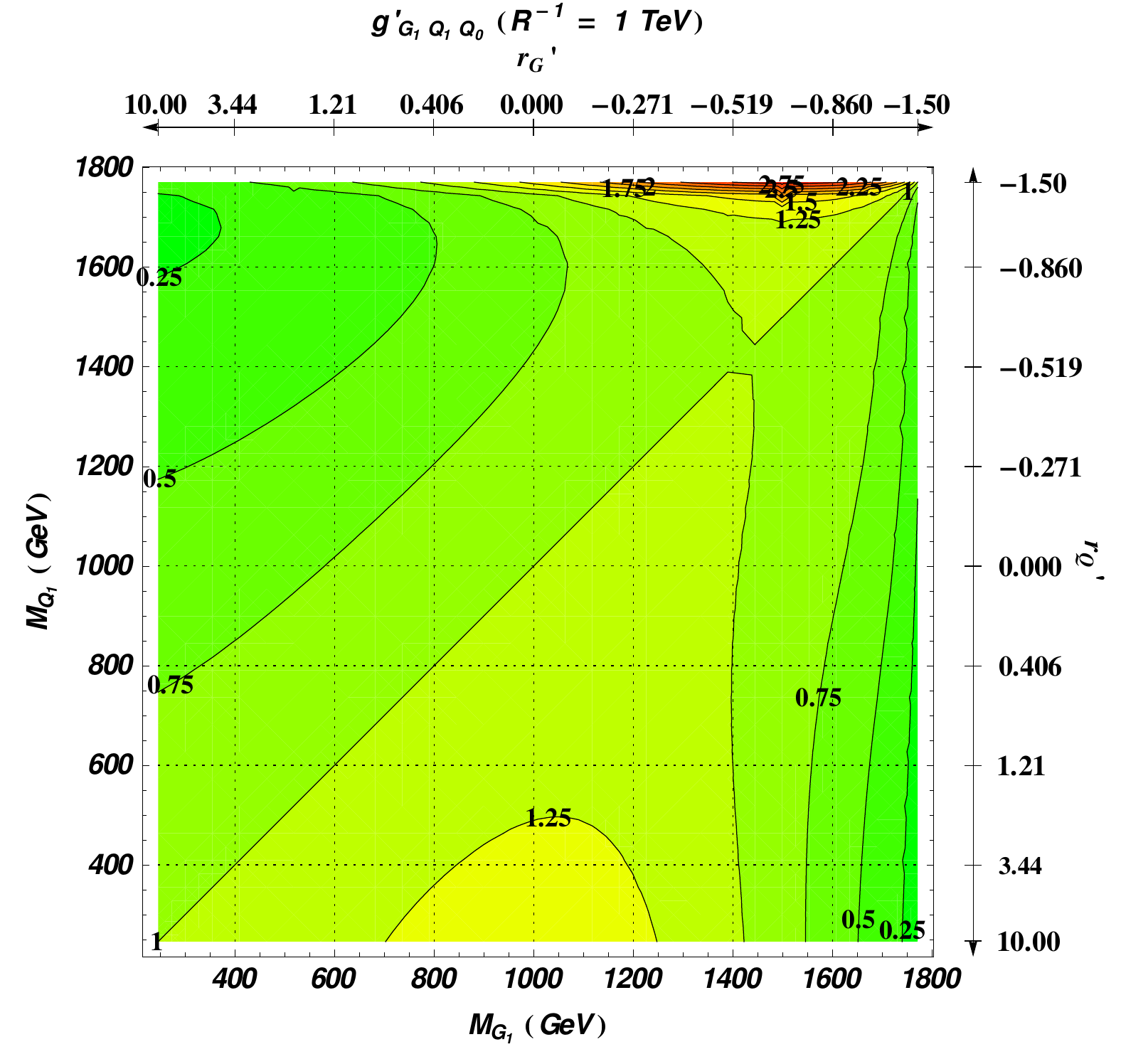}
\includegraphics[width=0.49\columnwidth , clip]{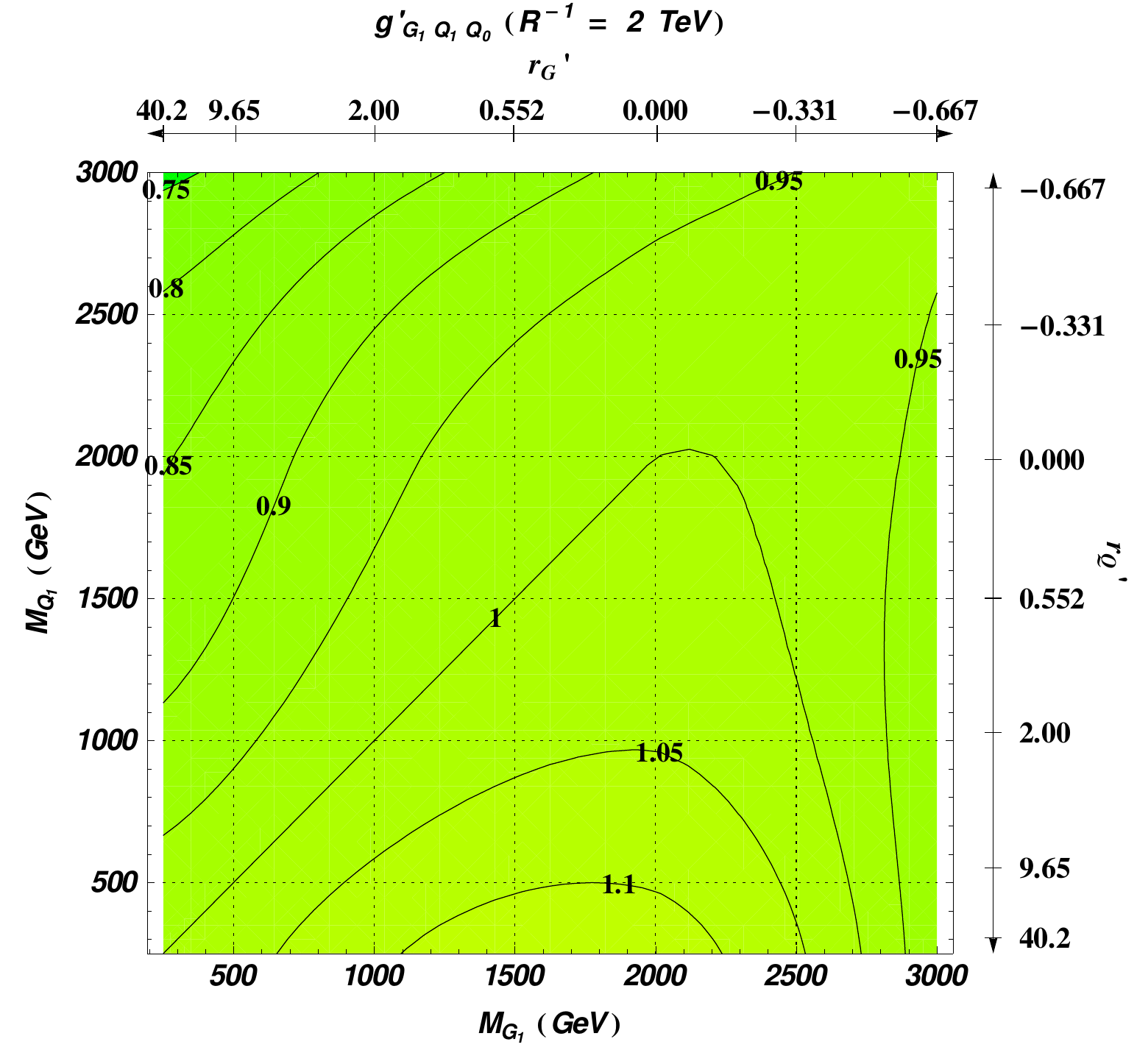}
\vskip 0.0in
\includegraphics[width=0.49\columnwidth , clip]{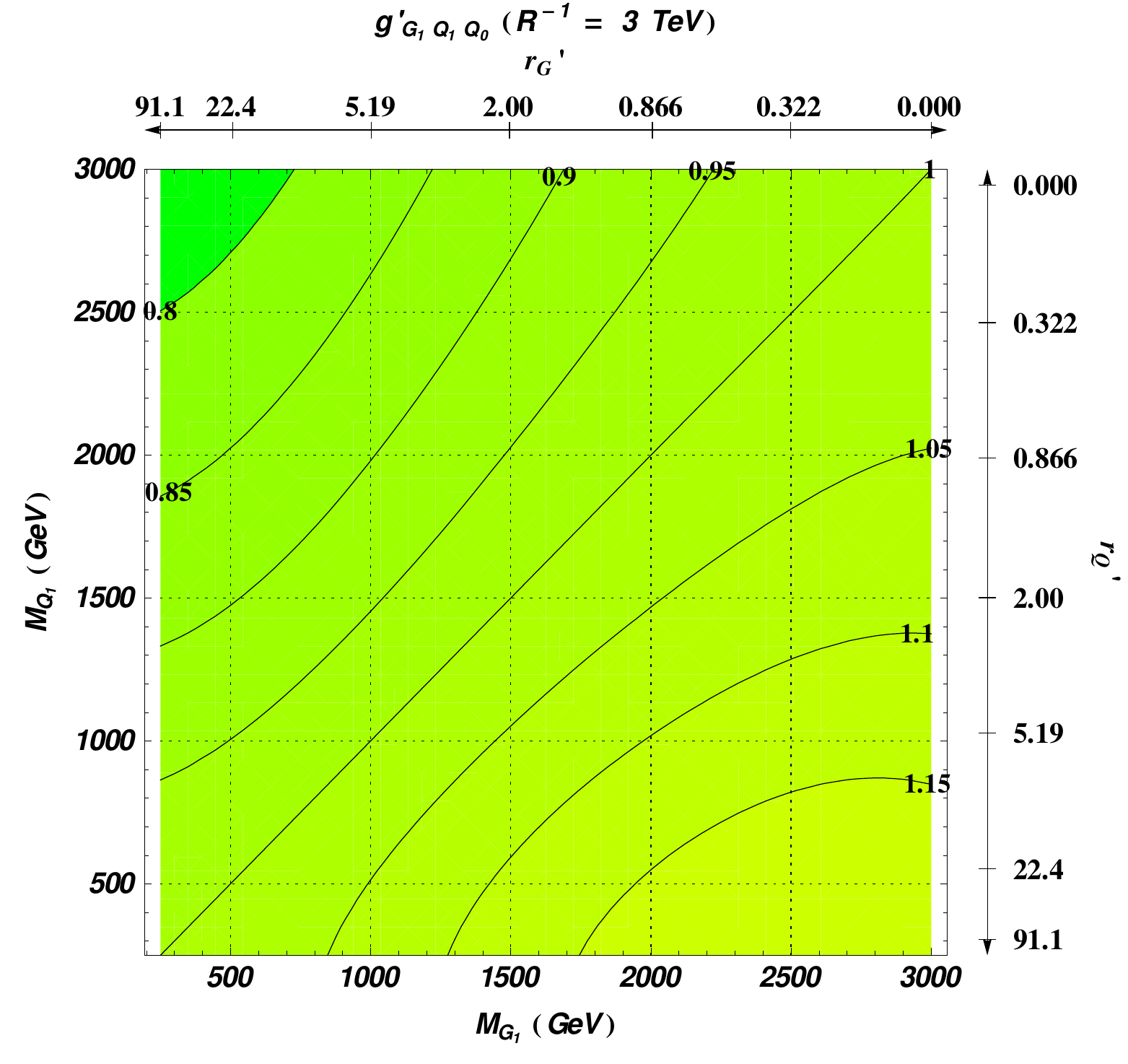}
%\hspace*{0.2cm}
\includegraphics[width=0.49\columnwidth  , clip]{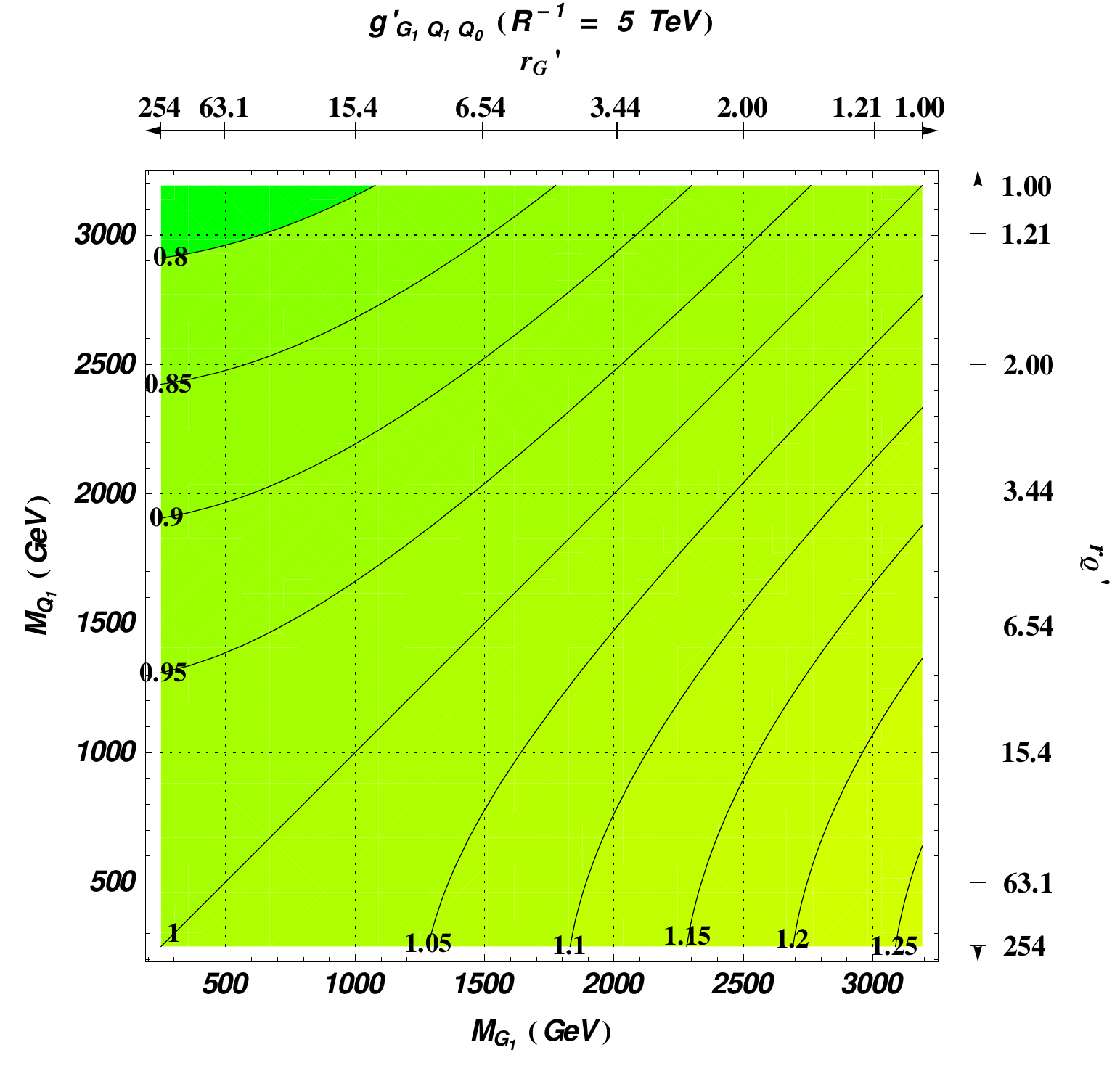}
\caption{Contours of the coupling deviation-factor $g'_{G_1Q_1Q_0}$ in the
{$M_{G_{(1)}}-M_{Q_{(1)}}$ (KK)} mass-plane for $\rinv=1$ TeV (top, left), 
$\rinv=2$ TeV (top, right), $\rinv={3}$ TeV (bottom, left) and 
$\rinv=5$ TeV (bottom, right).} 
\label{coup-mg-mq}
\end{figure}

To be convinced that such an approach would work, one has to demonstrate 
quantitatively that the value of $\rinv$ can be estimated reasonably 
correctly. There are prima facie evidence that such an estimate would be
unambiguous. This follows from the observation that neither the
$g'_{G_1 Q_1 Q_0}$ contours in figure \ref{coup-rg-rq} nor the same 
in figure \ref{coup-mg-mq} intersect each other.
In table \ref{tab:extract} we demonstrate the situation with some actual 
numbers for two different values of $g'_{G_1 Q_1 Q_0}$ which one might 
be able to extract from experiments. The number presented in the table 
are picked up directly from the contour-plots in figure~\ref{coup-mg-mq}.
Note that the values 0.85 and 1.1
that are chosen for $g'_{G_1 Q_1 Q_0}$ in table \ref{tab:extract} could 
result in $\sim 50\%$ deviations from the nominal values of the cross sections
(which {go} as $g{'}_{G_1 Q_1 Q_0}^4$) for the strong production 
modes at the LHC. This kind of a departure {can be} expected to be measured 
efficiently enough and thus can be used for further inferences. 
It is then informative to find from table \ref{tab:extract} that for an
experimentally estimated value of $g'_{G_1 Q_1 Q_0}$ and for a known
set of masses for the KK gluon and KK quarks, the value of $\rinv$ is pretty
distinct and thus can be estimated unambiguously.

\begin{table}[H]
\begin{center}
\begin{tabular}{||c | c |  c | c ||}
\hline
\hline
$g'_{G_1 Q_1 Q_0}$  & $R^{-1}$ (TeV) & ${M_{G_{(1)}}}$ (GeV) & ${M_{Q_{(1)}}}$ (GeV) \\
\hline
\hline
 & 2 & 835.1 & 2724.3  \\
0.85 & 3 & 840.9 & 2407.6  \\
 & 5 & 1246.1 & 2819.1 \\
\hline
 & 2 & 1820.0 & 500.0  \\
1.1 & 3 & 2019.5 & 1036.0  \\
 & 5 & 2121.7 & 989.6 \\
\hline
\hline
\end{tabular}
\caption{{(KK)-mass-values} for level `1' KK gluon/quark in nmUED for varying two representative
values of $g'_{G_1 Q_1 Q_0}$ and for varying $\rinv$. The choice of values for
$g'_{G_1 Q_1 Q_0}$ is motivated by figure \ref{coup-mg-mq}.}
\label{tab:extract}
\end{center}
\end{table}

%%%%%%%%%%%%%%%%%%%%%%%%%%%%%%%%%%%%%%%%%%
\section{Phenomenology at the LHC}
\label{sec:lhc}

In this section we discuss the cross sections of the  level `1' KK 
gluon ($G_1$) and quarks ($Q_1$) of the nmUED scenario produced via 
strong interaction at the LHC.
{Hereafter, we use simplified notations, $m_{G_{1}}$ and 
$m_{Q_{1}}$, to denote the physical masses of the level `1' KK gluon
and quarks, respectively.}
The patterns are explained by
relating them to the features of the scenario as discussed in detail in
sections \ref{sec:masses} and \ref{sec:couplings}. 
We then proceed to contrast the production-rates with the corresponding ones from 
mUED and SUSY. We also discuss at length the overall implications of such an nmUED scenario
whose signals can be faked by the latter two.

\subsection{Production cross sections for level `1' KK gluon and quarks}
\label{sec:cross-sections}

In figures \ref{fig:8tev-rates} and \ref{fig:14tev-rates} we present the 
cross sections for different final states for the 8 TeV and 14 TeV runs 
of the LHC, respectively, in the $r'_G-r'_Q$ plane. Results for generic 
final states like $G_1 G_1$, $G_1 Q_1$ and $Q_1 Q_1$ are laid out
in separate columns (from left to right) while separate rows are used 
to present the results for $\rinv=$ 1 TeV, 3 TeV and 5 TeV (from the 
top to the bottom). The final state indicated by $G_1 Q_1$ includes 
contributions from both $G_1 Q_1$ and $G_1 \bar{Q}_1$ while under 
$Q_1 Q_1$ we combine the rates from $Q_1 Q_1$, $Q_1 \bar{Q}_1$ and 
$\bar{Q}_1 \bar{Q}_1$. The rates include contributions from five 
flavours of level `1' KK quarks that correspond to five light SM quarks. 
For these states, as pointed out in sections \ref{sec:masses} and 
\ref{sec:couplings}, the role of $r'_Y$ is not significant 
except for some extremal cases, \emph{e.g.}, when $r'_Y \gg 1$,
for smaller $\rinv$.
Hence, we adopt a simplifying scheme where we set $r'_Q=r'_Y$ while
analyzing these excitations at the LHC.
Also, the contributions from both $SU(2)_W$-doublet and $SU(2)_W$-singlet 
varieties of KK quarks are included.  On the other side of the story, 
we have seen in section \ref{sec:masses} that the top quark sector 
turns out to be rather special thanks to crucial interplay of $r'_Q$, 
$r'_Y$ and $\rinv$ and to the possibility of significant level-mixing. 
This would render the phenomenology of the KK top quarks at the LHC rather 
rich. Given the intricacies involved, the analysis of this sector deserves a
dedicated study. This will be taken up in a future work.

The cross sections are calculated using MadGraph-5 \cite{Alwall:2011uj} 
in which the strongly interacting sector of the scenario is implemented 
through FeynRules \cite{Christensen:2008py} via its UFO (Universal 
FeynRules Output) \cite{Degrande:2011ua, deAquino:2011ub} interface. The mUED implementation
\cite{Datta:2010us} of CalcHEP \cite{Pukhov:2004ca} has been used for cross 
checks in appropriate 
limits and for some actual computation of cross sections in mUED. We used CTEQ6L 
\cite{Pumplin:2002vw} parametrization for the parton distribution 
function. The factorization/renormalization scale is fixed at the sum
of the masses of the final-state particles.
{In the remaining part of this work, we refer only to the physical masses
$m_{Q_{1}}$. These are the degenerate mass-eigenvalues obtained by diagonalizing the 
KK quark mass-matrix in the presence of brane-localized Yukawa terms and practically
same as the KK masses for the light quark flavours.} 

Some features common to both figures \ref{fig:8tev-rates}
and \ref{fig:14tev-rates} are as follows:
\begin{itemize}
\item the maximum value of the mass for the level `1' KK gluon and quarks 
      considered for $\sqrt{s}=8$ TeV (14 TeV) run of the LHC is 2 TeV 
      (3 TeV) which happens to be the tentative (perhaps, optimistic) 
      reach of LHC running at this center of mass energy.
      The conservative lower limit of the masses that has gone into
      the analysis is 500 GeV,
\item for given values of $\rinv$, the various ranges of $r'_G$ and $r'_Q$ in different rows
      ensure $m_{G_1}$ and $m_{Q_1}$ in the above-mentioned ranges,
\item to capture cross-sections varying over orders of magnitude, the
      contours are drawn after taking {the} logarithm (to base 10) {of the cross sections}.
      We, thus, encounter negative-valued contours in these figures,
\item for final states containing one or more level `1' KK quark (the second 
      and the third columns), the contour values, ({\emph{i.e.}}, the cross 
      sections) rise along the diagonal connecting the bottom-left and {the}
      top-right corners of the plots. This can be understood in terms of 
      decreasing $m_{G_1}$ and $m_{Q_1}$ as both $r'_G$ and $r'_Q$
      increase in that direction,
\item the {variation in the $G_1 G_1$ production (the first column) has
      a curious trend when compared} with the final-states having $Q_1$. The parallel, 
      vertical stripes (except in some region with $r'_G, \, r'_Q <0$ {only} for 
      low $\rinv$ ($\sim 1$ TeV)) imply that the cross-sections almost do 
      not vary with $r'_Q$. This means they are insensitive to variations 
      in {$m_{Q_{1}}$}. This is because the event rate for this final state 
      {is} dominated by the $s$-channel (gluon-fusion) subprocess where
      $Q_1$ plays no role unlike in the $t$-channel where the latter can 
      appear as a propagator. Hence, we see a gradual, steady increase 
      in rates only with increasing $r'_G$, \emph{i.e.}, with decreasing 
      $m_{G_1}$ which is quite expected.
\item the local dependence of the $G_1 G_1$ rate on $r'_Q$, for $r'_G, 
      \, r'_Q <0$ and $\rinv \sim 1$ TeV, shows a different trend. 
      In this region (from the deep blue to the white passing through 
      the light blue region), the rate grows in a direction of increasing
      $m_{Q_1}$ which is somewhat not so intuitive. It is instructive to observe 
      that for such a region, $m_{G_1}$ also turns out to be relatively 
      heavy (since, $r'_G<0$). Our probe into the phenomenon
      revealed that over this region the $t$-channel contribution becomes 
      important\footnote{Presumably, this happens since a larger $m_{G_1}$ 
      requires a larger $\sqrt{\hat{s}}$, in turn resulting in a lower 
      partonic flux for the gluon in the protons that ultimately results 
      in a suppressed contribution from the $s$-channel.} 
      and the relative values of $m_{G_1}$ and $m_{Q_1}$ are such that 
      a perceptible destructive interference takes place between $s$ 
      and  $t$ channels. Note also that with $g'_{G_1 Q_1 Q_0}$ getting
      extremally large over this region (see figure \ref{coup-rg-rq}) of 
      the parameter space, the overall situation gets further compounded,
\item the explanation holds for any final state that receives
      significant contribution{s} from subprocesses initiated by gluon(s).
      Thus, it is not unexpected that rates for $G_1 Q_1$ final state 
      show a similar behaviour in the said region of the parameter space
      while the same for the level `1' quark-pair final state, dominated
      by $Q_1 Q_1$ (which is not gluon-induced), {though rich in feature},
      do not show such a trend very clearly, 
\item for the $Q_1$-pair final state, one finds that in the region of 
      low $r'_Q$ ($<0$) the contours of larger cross sections reappear
      as one goes further down in $r'_Q$. This seems to be a result
      of extremally large value of $g'_{G_1 Q_1 Q_0}$ which can be
      understood from the region shaded in red in the right plot of
      figure \ref{coup-rg-rq}. Closer inspection reveals that the small, 
      yellow contour at the bottom of these plots exactly correspond to 
      the region of parameter space shaded in red in 
      figure \ref{coup-rg-rq}. In this region, naively, the 
      boost in cross section can be up to a factor $g'{^4_{G_1 Q_1 Q_0}}$
      which turns out to be $\approx 30$,
\item as we go from $G_1 G_1$ production to 
      $Q_1 Q_1$ production passing through $G_1 Q_1$ production the
      contours get flattened up in an anti-clockwise direction.
      This is easy to understand in terms of an increased dependence 
      of the rates on {$m_{Q_{1}}$} and hence, on $r'_Q$,
\item {it} may be noted that in the top panel of both figures 
      \ref{fig:8tev-rates} and \ref{fig:14tev-rates} (with $\rinv=1$ TeV) 
      the cross sections are not actually defined along the straight line
      with $r'_Q=0$. This is because some elements of the matrix in
      equation (\ref{eqn:Vmatrix1}) which enter the involved couplings
      for these final states are not defined at $r'_Q=0${,}
\item negative values of $r'_G$ and $r'_Q$ 
      are not considered for $\rinv$=3 TeV and 5 TeV cases since these take
      {$m_{G_{1}}$ and $m_{Q_{1}}$} far above the LHC reach. Thus, as we do not 
      enter the ``exotic'' part of the parameter space (with both 
      $r'_G, \, r'_Q <0$), we do not see any special variation in the 
      contour-patterns at lower values of $r'_G$ and $r'_Q${.}
\end{itemize}
The only major difference of a generic nature that we see between the
results presented in figures \ref{fig:8tev-rates} ($\sqrt{s}=8$ TeV) and 
\ref{fig:14tev-rates} ($\sqrt{s}=14$ TeV) is that for similar values of 
$r'_G$ and $r'_Q$, \emph{i.e.}, for similar values of {$m_{G_{1}}$ and 
$m_{Q_{1}}$} {for} a given $\rinv$, the rates are higher for the 14 TeV 
run, as expected.   

%%%%%%%%%%%%%%%%%%%%%%%%%%%%%%%%%%%%%%%%%%

\begin{figure}[H]
\centering
%\vspace*{4in} \hspace*{-3.9in}
\includegraphics[width=\columnwidth, clip]{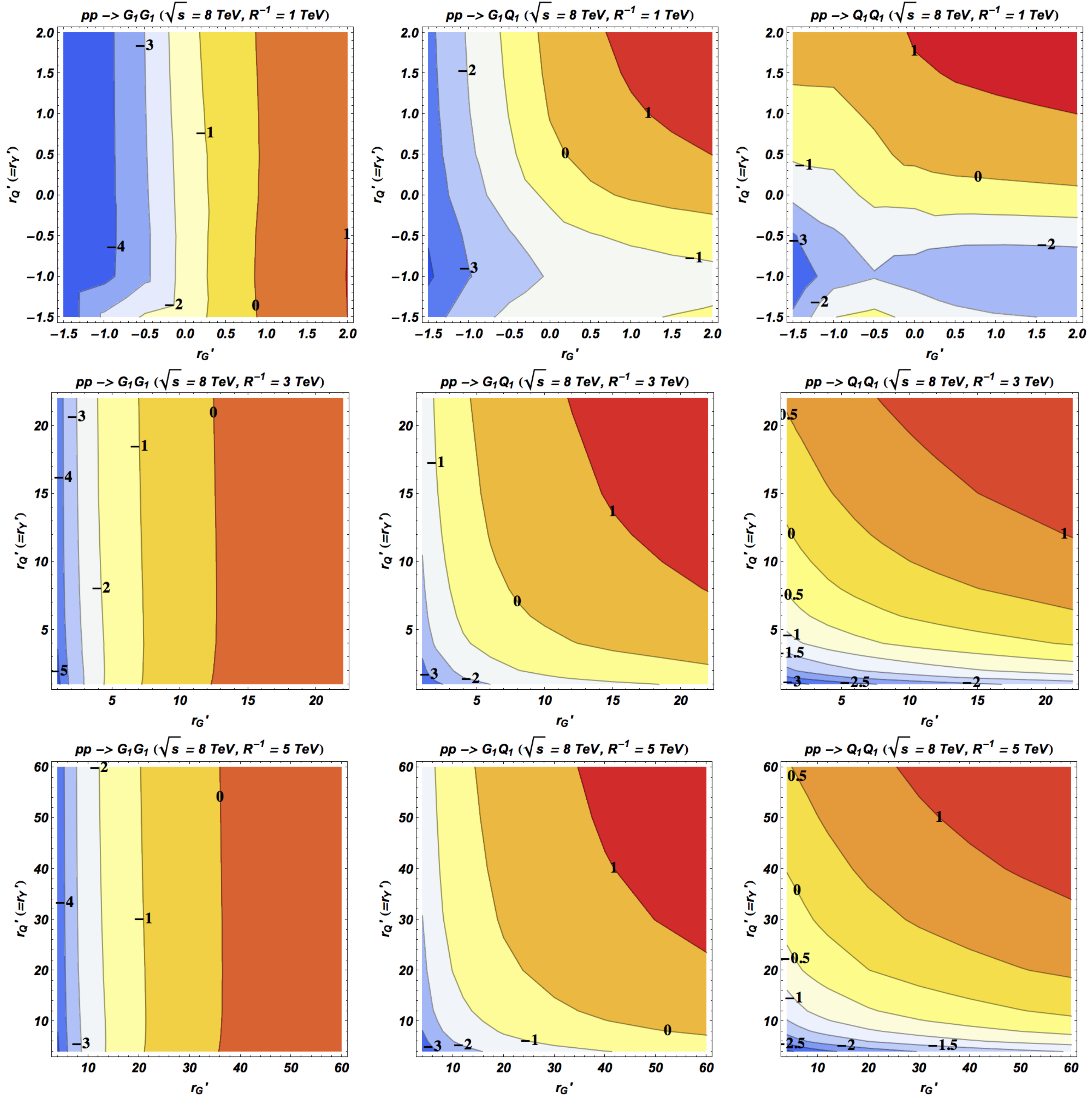}
%\vskip 20pt
%\hspace*{1cm}
\caption{Log-valued (to base 10) cross section (in pb) contours
for different final states
at the LHC for $\sqrt{s}=8$ TeV in the $r'_G-r'_Q$ plane with 
$\rinv$ as a parameter. $\rinv$ varies across the rows while each
column specifies a particular final state. CTEQ6L parametrization 
is used for the parton distribution function. 
The factorization/renormalization scale is fixed at the sum of the 
masses of the two final-state particles. To find the conventions
adopted in clubbing individual final states into generic ones, please
refer to the text.}
\label{fig:8tev-rates}
\end{figure}

\newpage
\begin{figure}[H]
\centering
%\vspace*{4in} \hspace*{-3.9in}
\includegraphics[width=\columnwidth, clip]{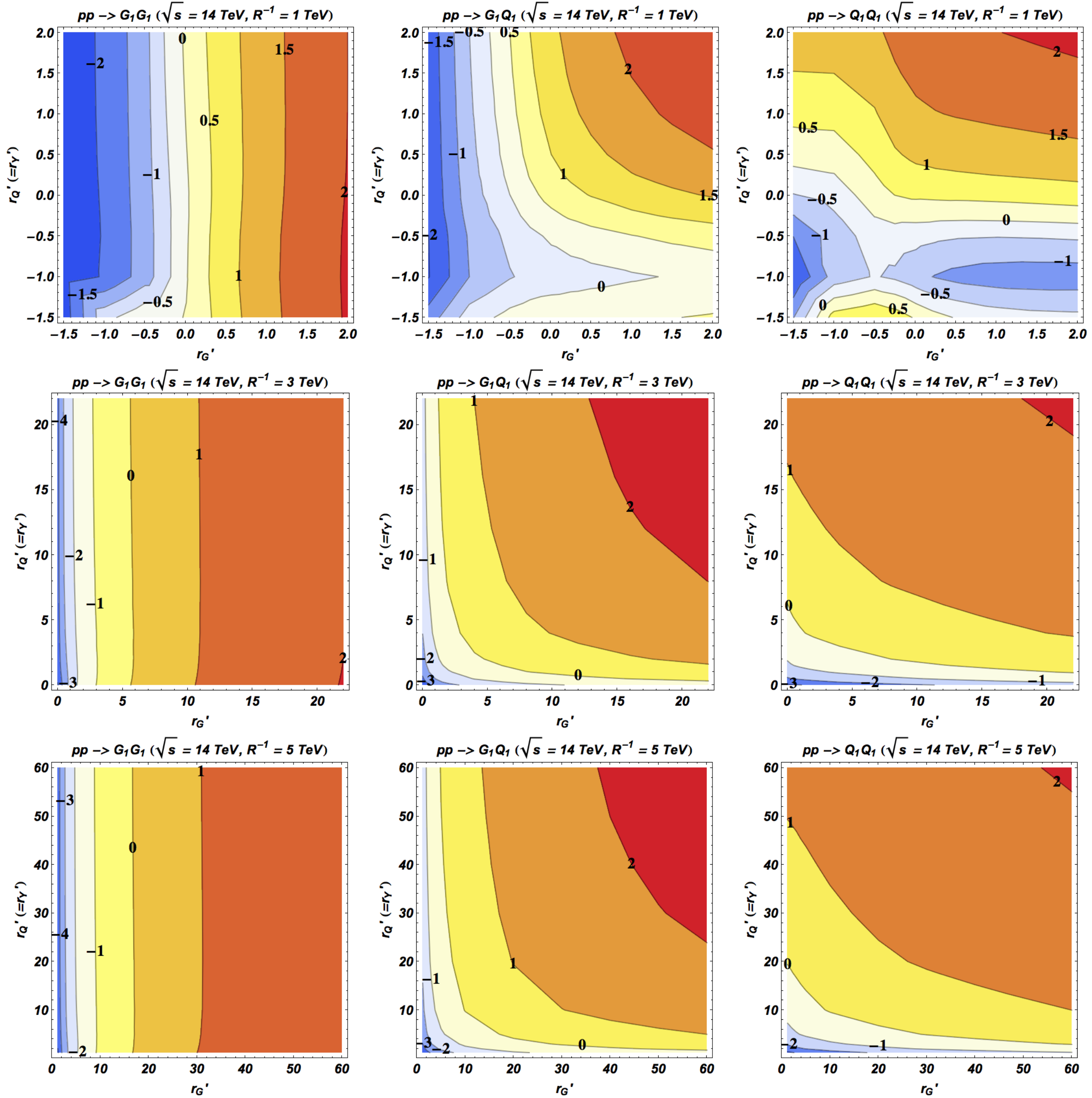}
%\vskip 20pt
%\hspace*{1cm}
\caption{Same as in figure \ref{fig:8tev-rates} but for
$\sqrt{s}=14$ TeV.}
\label{fig:14tev-rates}
\end{figure}

%%%%%%%%%%%%%%%%%%%%%%%%%%%%%%%%%%%%%%%%%
\subsection{mUED vs nmUED vs SUSY}
In this subsection we take up the interesting possibility of mUED and nmUED
faking each other and faking SUSY as well. This is reminiscent of the 
possibility of UED faking SUSY \cite{Cheng:2002ab} 
where one talks about a situation in which the final state masses happen 
to {be consistent with} a mUED-like spectrum \cite{Datta:2005vx, Smillie:2005ar}. 
There, SUSY being a less constrained scenario, 
this is the natural set up to study its faking by mUED. Thus, with more free 
parameters in the scenario, nmUED may enjoy a {more} direct parallel to SUSY 
when being compared with mUED.\footnote{
Going one step further, it may be said 
that faking between UED and SUSY tend to get more complete~\cite{Datta:2005zs, Datta:2005vx} with an nmUED-type 
scenario for which the mass{es} of the KK excitation{s} may take almost any arbitrary 
values. In this sense, it may be interesting
to note the apparent contrast in the naming schemes for scenarios in SUSY and those
{involving a} UED framework. In the case of SUSY, the minimal version is the least constrained 
one (with too many free parameters) while the same for UED is the one which is its
most constrained incarnation with only two (three, with level `0' Higgs mass parameter)
parameter{s}.}

In table \ref{mued-vs-nmued} we compare the cross sections for the production
of level `1' KK gluon and quarks as obtained from the mUED scenario and the nmUED 
version we are studying in this work at the LHC with $\sqrt{s}=14$ TeV. 
Assuming that the ballpark values of the
masses of these excitations could be anticipated once a positive signal is found 
at the LHC, we fix {these} masses to carry out the analysis. 

The reference values for the masses employed in table \ref{mued-vs-nmued} 
are $m_{G_1}=1220$ GeV, 
$m_{Q^D_1}=1154$ GeV and $m_{Q^S_1}=1133$ GeV. These are obtained in the mUED 
scenario by 
setting $\rinv=1000$ GeV and $\Lambda R =10$. For the nmUED scenario, we
require the level `1' gluon mass to be the same as in the case of mUED while
for the doublet and singlet KK quarks we take a common value which is almost equal
to the singlet one in mUED. Note that, in the absence of radiative corrections,
the masses of the doublet and singlet KK quarks are the same in the nmUED 
scenario under consideration and both are determined by the brane-localized 
parameter $r'_Q$.
Such an nmUED spectrum is generated for different $\rinv$ by suitably tuning
the brane-localized parameters $r'_G$ and $r'_Q$. 

\emph{A priori}, a comparison of cross sections from the two scenarios having similar
spectrum assumes a special significance since the brane-localized parameters,
$r'_G$ and $r'_Q$, not only control the KK masses but also affect their couplings.
These are discussed in sections \ref{sec:masses} and \ref{sec:couplings} with 
illustrations (see figures \ref{coup-rg-rq} and \ref{coup-mg-mq}){.}
It can be gleaned from table \ref{mued-vs-nmued} that except for the case 
where {$\rinv_{mUED} > \rinv_{nmUED}$} with $\rinv_{nmUED}= 700$ GeV and leaving
out the $G_1$-pair final state, the cross sections for the rest are within 
$\sim 10$\% of the corresponding mUED values. For these cases, the reason of such 
a closeness in cross sections can be understood in terms of the small deviation of 
the strong coupling from the mUED case which is quantified by $g'_{G_1 Q_1 Q_0}$
and indicated in column 2. 
{The smallness of the deviation in $g'_{G_1 Q_1 Q_0}$ is ensured by the
requirement of near-identical values of $r'_G$ and $r'_Q$ in nmUED that reproduce
the characteristic splitting between the masses of the KK gluon and the quarks
in mUED.}

On the other hand, the case for the $G_1$-pair production is somewhat interesting.
There, the cross sections are insensitive to variation in $g'_{G_1 Q_1 Q_0}$ 
in contrast to what we see in case of other final states as we move on from 
$\rinv$=700 GeV. This may be attributed to the fact that the modified coupling 
given by $g'_{G_1 Q_1 Q_0}$ only appears in the $t$-channel while the process 
$pp \to G_1 G_1$ gets dominant contribution from the $s$-channel. 
Moreover, unlike the previous cases, here, a marked difference is noticed
between the cross sections for the mUED and the nmUED scenarios with the nmUED 
cross section ($\sim 0.17$ pb) being $\sim 20\%$ smaller than the corresponding
mUED value ($\sim 0.22$ pb). The reason for this can be traced back to the particular
chiral structure of the interaction vertex originating from the action in 
equation (\ref{eqn:s-quark-int}) that contains the elements $v$-s of the $V$ matrices 
(see equations (\ref{eqn:Vmatrix1}),\,(\ref{mUEDlimitingbiunitarymatrix}) and (\ref{eqn:Vmatrix2})).

\begin{table}[H]
\begin{center}
{\small
\begin{tabular}{||c|c||c|c|c|c|c|c||}
\hline
\hline
\multicolumn{2}{|c|}{mUED Parameters} & \multicolumn{6}{|c|}{$\rinv= 1000$ GeV, $\Lambda R = 10$} \\
\hline
\multicolumn{2}{|c|}{mUED/SUSY Mass (in GeV)}   & \multicolumn{6}{|c|}{$m_{{G_1}/\tilde{g}}$= 1220 
                                       $\quad m_{{Q^D_1},\tilde{q}_L}$= 1154
                                       $\quad m_{{Q^S_1},\tilde{q}_R}$= 1133} \\
\hline
\hline
\multicolumn{2}{|c|}{}  & \multicolumn{6}{|c|}{} \\
\multicolumn{2}{|c|}{}     & \multicolumn{6}{|c|}{Cross sections (in pb)} \\
\multicolumn{2}{|c|}{}  & \multicolumn{6}{|c|}{} \\
\hline
\multicolumn{2}{|c|}{Final states} &  $G_1 G_1$              &   $G_1 Q_1$        &   $G_1 \bar{Q}_1$   &   $Q_1 Q_1$   
          &  $\bar{Q}_1 \bar{Q}_1$  &   $Q_1 \bar{Q}_1$  \\
\hline
\hline
\multicolumn{2}{|c|}{}  & & & & & & \\
\multicolumn{2}{|c|}{mUED} & 0.216 & 1.250 & 0.082 & 1.132 & 0.009 & 0.403 \\
\multicolumn{2}{|c|}{}  & & & & & & \\
\hline
\hline
 & & & & & & & \\
 & $\rinv$=700 GeV & & & & & &  \\
 & $r'_G=-1.34$   & & & & & &  \\
       & $r'_Q=r'_Y=-0.90$   & & & & & &  \\
 & $g'_{G_1 Q_1 Q_0}$=0.627   & 0.178 & 0.503 & 0.032 & 0.177 & 0.001 & 0.173 \\
 & & & & & & & \\
\cline{2-8}
 & & & & & & & \\
 & $\rinv$=1000 GeV & & & & & &  \\
 & $r'_G=-0.30$   & & & & & &  \\
       & $r'_Q=r'_Y=-0.19$   & & & & & &  \\
 & $g'_{G_1 Q_1 Q_0}$=1.035   & 0.172 & 1.349 & 0.085 & 1.277 & 0.009 & 0.432 \\
 & & & & & & & \\
\cline{2-8}
 & & & & & & & \\
 & $\rinv$=1500 GeV & & & & & &  \\
nmUED & $r'_G$=0.37   & & & & & &  \\
 & $r'_Q$=$r'_Y=$0.54   & & & & & &  \\
 & $g'_{G_1 Q_1 Q_0}$=1.033   & 0.173 & 1.364 & 0.086 & 1.303 & 0.010 & 0.438 \\
 & & & & & & & \\
\cline{2-8}
 & & & & & & & \\
 & $\rinv$=2000 GeV & & & & & &  \\
 & $r'_G$=1.15   & & & & & &  \\
       & $r'_Q$=$r'_Y=$1.43   & & & & & &  \\
 & $g'_{G_1 Q_1 Q_0}$=1.026   & 0.171 & 1.336 & 0.084 & 1.262 & 0.009 & 0.427 \\
 & & & & & & & \\
\cline{2-8}
 & & & & & & & \\
 & $\rinv$=2500 GeV & & & & & &  \\
 & $r'_G$=2.13   & & & & & &  \\
       & $r'_Q$=$r'_Y=$2.56   & & & & & &  \\
 & $g'_{G_1 Q_1 Q_0}$=1.019   & 0.172 & 1.326 & 0.083 & 1.233 & 0.009 & 0.421 \\
 & & & & & & & \\
\hline
\hline
\multicolumn{2}{|c|}{}  & & & & & & \\
\multicolumn{2}{|c|}{SUSY (MSSM)} & 0.019 & 0.181 & 0.012 & 0.153 & 0.001 & 0.054 \\
\multicolumn{2}{|c|}{}  & & & & & & \\
\hline
\hline
\end{tabular}
}
\caption{Comparison of the cross sections in mUED, nmUED and SUSY (MSSM) scenarios for similar spectra
at the LHC with $\sqrt{s}=14$ TeV.
In mUED the spectrum is generated for a given $\rinv$ (1 TeV). In nmUED
matching spectra are generated by varying $\rinv$ and tuning the values of
$r'_G$ and $r'_Q$ simultaneously while keeping $r'_Y=r'_Q$. For SUSY, the masses of
the corresponding excitations (indicated clearly against the mass variables) are
tuned to similar values by varying the soft SUSY breaking parameters appropriately.
CTEQ6L parton distribution functions are used and the renormalization/factorization
scale is set at the sum of two final state masses.}
\label{mued-vs-nmued}
\end{center}
\end{table}

The differences in the cross sections, as we see from table \ref{mued-vs-nmued},
for the mUED and the nmUED scenarios, {are} not big enough for the LHC to signal a clear departure 
from one or the other of the two competing scenarios. Thus, it turns out that if 
a spectrum is compatible with the mUED scenario, it would not be easy to
rule out a non-minimal version of the UED solely based on such a study. Of course,
it may happen that other simultaneous studies involving the electroweak sector
could help distinguish between the two.

The last line in table \ref{mued-vs-nmued} shows the corresponding cross sections in a
SUSY scenario (based on Minimal Supersymmetric Standard Model (MSSM)). The level `1'
KK excitations of the UED scenarios are substituted by their counterparts in SUSY: the KK 
gluon by the gluino, the $SU(2)_W$-doublet quark by the left handed squark and the
$SU(2)_W$-singlet quark by the right handed squark. It is  well known that, for
identical mass spectra, UED production cross sections are {generically} larger than that for the
analogous SUSY processes (by roughly a factor between 7 and 10). 
This is partly related to the structure of the UED matrix elements 
and the extra helicity states that UED excitations possess when compared to an analogous final state
in SUSY. Even then, it is interesting to find that for $g'_{G_1 Q_1 Q_0} < 1$ (the first
entry for the nmUED case in table \ref{mued-vs-nmued}), cross sections in some of the final states could approach {the}
SUSY values. Thus, the total rate for strongly produced particles ceases to be a good
enough indicator for the underlying scenario. This brings the alleged faking to an almost 
complete level. 
Note that this kind of a possibility does not arise in mUED. This again highlights how the
correlation between masses and the couplings of the KK excitations in the nmUED scenario
could shape the phenomenological situation in an interesting and involved way.
%
%%%%%%%%%%%%%%%%%%%%%%%%%%%%%%%%%%%%%%%%%%%%%%%%%%%%%%%%%%%%%%%%%%%%%%%%%%%%%%%%%%%%%%%%%

\subsection{Decays of level 1 KK gluon, quarks and electroweak gauge bosons}
\label{section:decays}

In this section we discuss in brief the decay patterns of the level 1
KK gluon and quarks. When the KK gluon is heavier than the KK quarks 
(mutually degenerate for the lighter generation of the quarks), it 
decays to $qQ_1$ final state with 100\% branching fraction. Thus,
cascades are governed by the decay of the level 1 KK quarks which,
in turn, decay to level 1 electroweak (EW) gauge bosons, $W^\pm_1,Z_1$ 
and $B_1$ in two-body modes. On the other hand, for $m_{Q_1} > m_{G_1}$,
level 1 KK quark undergoes 2-body decays to KK gluon and 
$W^\pm_1,Z_1$ and $B_1$. The KK gluon, in turn, decays to SM quarks
and the above set of electroweak KK gauge bosons via 3-body modes.

We work with an electroweak sector at the first KK
level (comprising of the gauge bosons, the charged leptons and the neutrinos) 
which is reminiscent of mUED with corrected masses \cite{Cheng:2002iz}, 
that are essentially 
determined by $R^{-1}$. {This can be seen as a limit of an electroweak
sector in nmUED with vanishings BLTs.} This is in line with the main goal of the present 
work as we focus on the role of BLKTs in the strongly interacting sector only.
The resulting framework could thus be considered as a suitable benchmark 
(with only two BLKT parameters, $r_G^\prime$ and {$r_Q^\prime$}) for initiating 
a phenomenological analysis of the nmUED at the LHC. 
{The interaction vertex $q Q'_1 V_1$ ($V_1$ being the level `1' electroweak 
gauge boson) that takes part in electroweak decays of the level `1' KK quarks 
gets modified and follows from equation (\ref{deviation-factor}) with $r'_G \to 0$.}
Of course, more involved 
studies in scenarios having BLTs for the electroweak sector are highly 
warranted since such scenarios could emerge as perfect imposters of their 
popular SUSY counterparts.

With this assumption, $W^\pm_1$ and $Z_1$
always decay to leptonic modes, \emph{i.e.}, $W^\pm_1 \to \ell_1 \nu / 
\nu_1 \ell$ and $Z_1 \to \ell_1 \ell , \nu_1 \nu$. $B_1$ is the lightest
KK particle (LKP) and is stable. The only requirement to ensure these in nmUED
is to set $r_G^\prime$ and {$r_Q^\prime$} in a way such that $ m_{G_1}$ and 
$m_{Q_1}$ do not become lighter than these electroweak 
bosons. This necessarily constrains the ranges of $r_G^\prime$ and 
{$r_Q^\prime$} that such a framework can take.

\vspace{0.25cm}
\begin{figure}[h]
\centering
%\includegraphics[width=0.60 \columnwidth, clip]{u1width.pdf}
%\hspace{-3.35cm}
%%\hskip -100pt
%\includegraphics[width=0.60 \columnwidth, clip]{d1width.pdf}
%\includegraphics[height=1in,width=1in,angle=0]{u1width}
\includegraphics[scale=0.60,angle=0]{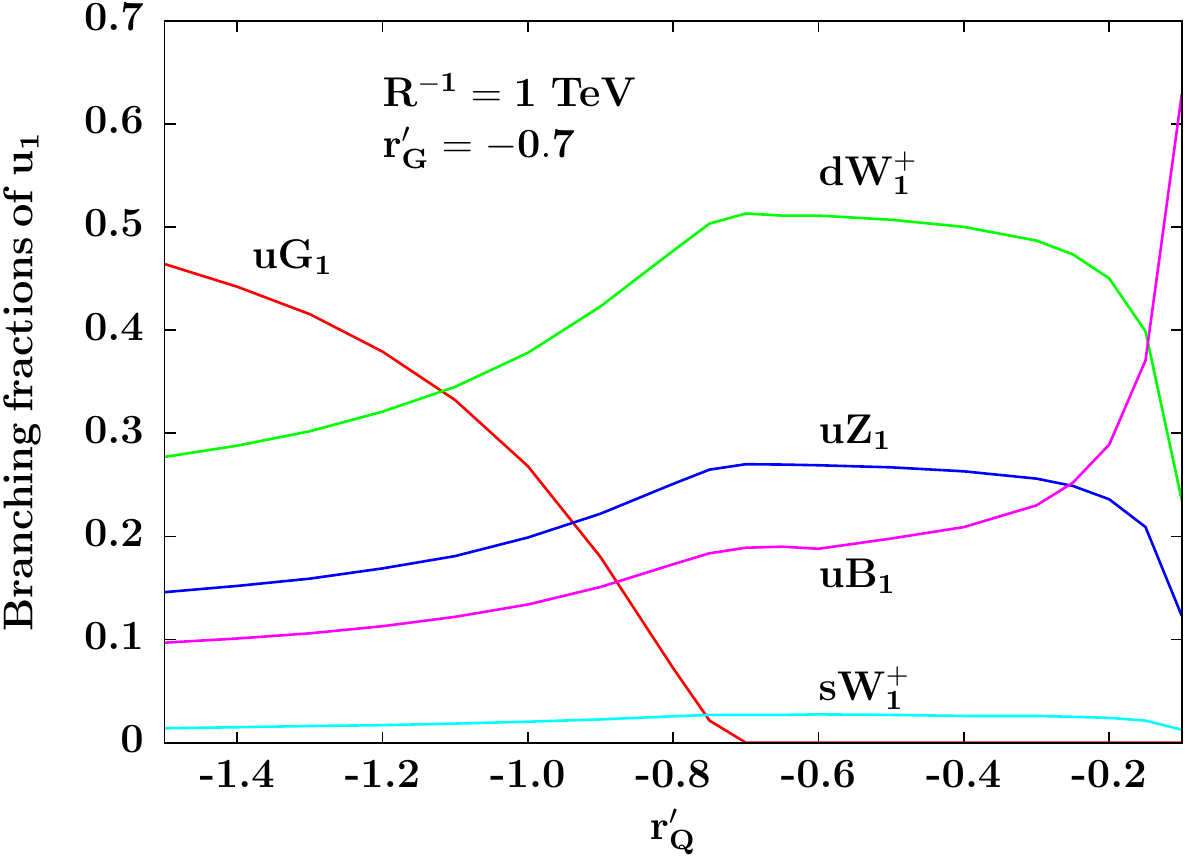}
\hskip 20pt
\includegraphics[scale=0.60,angle=0]{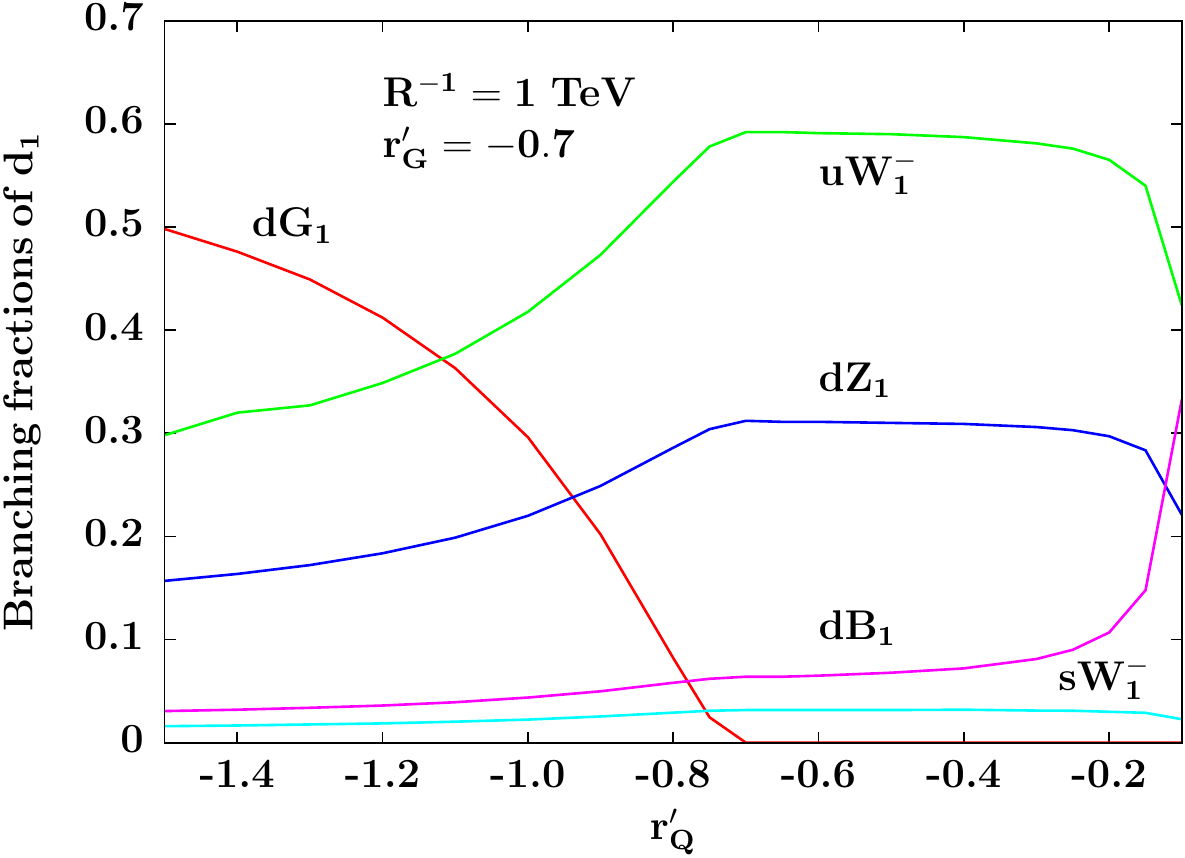}
%\includegraphics[scale=1.5]{u1width.pdf}
%\includegraphics[scale=0.5,angle=0]{d1width.pdf}
%\vspace{-8.00cm}
\caption{Branching fractions of level 1 \emph{up} (left) and 
\emph{down} (right) type KK quarks as functions of $r'_Q$ for 
$R^{-1}=1$ TeV. The level 1 gluon mass is taken to be 1.52 TeV 
which corresponds to $r'_G= -0.7$.} 
\label{fig:branchings}
\end{figure}

In figure \ref{fig:branchings} we present the branching fractions of the
up and down type KK quarks as functions of $r'_Q$ and for fixed values of
$R^{-1}$ (1 TeV) and $r'_G$ (-0.7). Each plot covers a range in {$r'_Q$} for
which both hierarchies between $m_{G_1}$ and $m_{Q_1}$ are realized.
Note that the decay widths (and the branching fractions) of the two
mass eigenstates for each type of KK quark are very similar since
they result from nearly maximal mixings of the weak eigenstates {(see equation \ref{eqn:Vmatrix1})}.
It is clear that for
$m_{{u_1/d_1}} > m_{G_1}$, the KK quarks may dominantly decay into
KK gluon with branching fractions reaching up to 50\% before
dropping quickly as quark mass increases.
For a reverse hierarchy, the KK quarks only have 2-body electroweak 
decays. Among these, decays to Cabibbo-enhanced $W^\pm_1$ dominate
followed by decays to $Z_1$, $B_1$ and Cabibbo-suppressed
$W^\pm_1$ modes. The difference between 
electroweak branching fractions of the $u$ and the $d$-type KK quarks 
stems solely from the difference in their hypercharges which
only affects their decay widths to $B_1$. It can be seen from
figure \ref{fig:branchings} that the peak branching fraction
to $W^\pm_1$ could be between 50\% and 60\% for the $u$ and the 
$d$-type KK quarks, respectively. The average branching fraction
to $Z_1$ is found to be between 20\% and 30\% in the range of $r'_Q$ where
electroweak decays dominate for the two species of quarks. Branching to
$B_1$ tends to remain at around 20\% (10\%) or less for the $u$-type
($d$-type) quark before it shoots up as the quarks become lighter
(from left to right)
and the splitting between them and $W_1^\pm$ and $Z_1$ become smaller. 
However, since the EW gauge boson masses are solely determined by
$R^{-1}$, all three EW decay modes remain healthy over entire
range of $r'_Q$.

Note that the branching fractions of the KK quarks to the stable 
LKP ($B_1$) is on the lower side and they hardly dominate (except
in extreme corners of the parameter space). 
This is in
stark contrast to mUED (or, for that matter, SUSY scenarios) where the right chiral 
level 1 KK quarks (right chiral squarks) decay almost 100\% of the time 
to $B_1$ LKP (bino-like LSP) when their strong decays to 
level 1 KK gluon (gluino) are closed. Later, we will see that 
this can have major implications for the relative rates in different final
states when compared to contending scenarios.

As mentioned earlier, for $m_{G_1} < m_{Q_1}$, KK gluon decays to level 1 EW gauge 
bosons ($W_1^\pm$, $Z_1$ and $B_1$) in three-body modes via off-shell 
KK quarks. The variations of these branching fractions with respect to 
$r'_Q$ (or $r'_G$) are expected to be flat. This is because $r'_Q$ ($r'_G$) appears in the 
primary vertex of these decays and in the propagator (through the KK 
quark mass) and these two affect all three body decay modes in a similar way.
The branching ratios are governed by the secondary vertex and thus 
follow the same pattern as in the decay of KK quarks, \emph{i.e.}, 
branching to $W_1^\pm$ dominates over the other two while the same to 
$B_1$ is the least favoured.\footnote{However, there may be a situation 
in nmUED when splitting between $m_{G_1}$ and $m_{W_1^\pm/Z_1}$ drops 
critically resulting in an enhanced branching to $B_1$.}

\subsection{Exclusion limits}

It is instructive to have a look at the current LHC data and understand
to what extent they may constrain an nmUED scenario of the present kind. 
In absence of a complete
implementation of the scenario in an event generator, we limit ourselves
to a parton level analysis which would, for example, give a ballpark estimate of the
exclusion limit for $R^{-1}$ under a reasonable set of 
assumptions.\footnote{Note that thorough simulation-studies for even the
mUED scenario are not yet existing in the literature.} In principle,
constraints can be derived on any subspace of the 3-dimensional space
spanning over $R^{-1}-r'_G-r'_Q$.

Here, we take up a recent ATLAS analysis \cite{:2012rz} of the final state 
with jets plus missing energy (with vetoed leptons) at $\sqrt{s}=7$ TeV
and integrated luminosity of 4.7 fb$^{-1}$.
We refer to the exclusion they report for equal mass gluino and squarks 
in the CMSSM scenario which is 1360 GeV. It must be pointed
out that a straight-forward comparison with the experimental data 
ultimately requires a 
thorough simulation (including the detector effects) of the nmUED scenario 
under consideration which should wait for a full implementation of the 
same in an event generator like MadEvent and/or others. Nonetheless, using 
the information we gathered in the last subsection, we can reasonably 
attempt to translate the above ATLAS bound to ballpark constraints on 
the nmUED scenario.

Towards this we find the value of cross section times branching fraction
(before cuts) for $m_{\tilde{q}} = {m_{\tilde{g}}} = 1360$ GeV using the
similar set of CMSSM parameters as in the ATLAS study. 
In the absence of
a complete simulation (where one would be able to employ {kinematic} cuts), 
we rely on this number and treat the same as the upper bound
on the cross section times branching fraction. 
The task is then to find the bound on the masses and/or the {parameters} of the
nmUED scenario that satisfies this constraint.

To carry out the analysis, we break the same up in three 
distinct regions in the nmUED parameter space having 
$m_{Q_1}>m_{G_1}$, $m_{Q_1} = m_{G_1}$ and $m_{Q_1}<m_{G_1}$. 
Since we are not in a position to use kinematic cuts employed
in the ATLAS analysis, the minimum requirement for being able to 
compare the nmUED results with the ATLAS study is to ensure the 
the kind of nmUED-spectra that result in hard enough jets and missing 
energy so that the ATLAS acceptances/efficiencies 
would hold safely.\footnote{The spectra for
this analysis are so chosen that for unequal masses for $Q_1$
and $G_1$, the mutual splitting between them as well as the 
splitting between the lighter one between $Q_1$ and $G_1$ and 
the LKP is around 200 GeV. This would ensure (in absence of a
full-fledged simulation with detector effects) jets from 
both primary and secondary cascades and the missing transverse
energy to be hard enough to pass strong ATLAS cuts.}.
Thus, the constraints we
obtain for the nmUED scenario could only be conservative and can 
be improved with help of a dedicated simulation. It is found that 
$R^{-1} < 950$ GeV could be ruled out for $m_{Q_1}<m_{G_1}$ while, 
for $m_{G_1}<m_{Q_1}$, the exclusion could at best be up to 900 
GeV.\footnote{Note that these bounds on $R^{-1}$ are insensitive 
to the values of $r'_Q$ and $r'_G$ as long as the spectral 
splittings demanded are satisfied. Qualitatively, this can be 
termed as the most stringent constraint that could be put on the 
three-dimension nmUED parameter space considered here. One
may like to take note of the anomalous region of a terminally large 
{negative} $r'_Q$ with large $m_{Q_1}$ for which the couplings become very
strong and could over-compensate for the suppression in the
cross section due to large $m_{Q_1}$. In this region, perhaps, a larger
value of $R^{-1}$ could be ruled out.} For $m_{G_1} \simeq m_{Q_1}$, the
lower bound can be as high as 1.1 TeV. However, in that case, sensitivity
would be higher in the signal region with not too many hard jets since
strong 2-body decays are phase-space suppressed. 

{In any case, we find that the bounds are degraded for nmUED when
compared to CMSSM. This is not unexpected because of lower yield in 
$jets+\slashed E_T$ channel for nmUED. On the other hand, similar 
constraints on CMSSM are expected to be weaker from the analysis of 
leptonic final states while
the same for nmUED would yield a more stringent bound.}

\subsection{The case for 14 TeV LHC}

In this subsection we discuss in brief the pattern of yields
in various multi-jet, multi-lepton final states accompanied
by large amount of missing transverse energy. The reference
values chosen for this discussion are $R^{-1}=1$ TeV and 
$r'_G=-0.7$ which are the same as in \ref{section:decays}.
In table \ref{table:predictions} we present the expected uncut 
yields (in fb) for
these final states as $r'_Q$ varies. For the second and
the third columns, $m_{Q_1}<m_{G_1}$ while for the fifth and the 
sixth columns, $m_{Q_1}>m_{G_1}$. For the fourth column  
$r'_Q=r'_G$ and hence $m_{Q_1}=m_{G_1}$. To highlight the contrast,
in the last three column we present
the corresponding numbers for the mUED cases where the scenarios
are solely determined by $R^{-1}$, for all practical purposes.

It is seen from table \ref{table:predictions} that yields for
all the final states decrease as $r'_Q$ decreases except for
$r'_Q=-1.5$ when the same increases suddenly. The latter can be 
understood in terms of an abrupt increase (up to three-fold) 
in the modified strong
coupling close to the boundary of the theoretically allowed nmUED
parameter space {(see figure \ref{coup-rg-rq})}.
The increase in the coupling strength, in fact, 
(over-)compensates for
the lowering of the strong production cross sections as $m_{Q_1}$ 
increases with decreasing $r'_Q$. The drop in the yields over
the range $-0.1> r'_Q > -1$ is attributed to the increase in
$m_{Q_1}$ when the increase in strong coupling strength is limited
to around 20\%. Note also the sharp variation of the yields for
all the final states when going from $r'_Q =-0.1$ to $r'_Q =-0.5$.
This is mainly due to a sharper rise in $m_{Q_1}$ (by $~300$ GeV) 
when compared to the columns to follow (for which the rises are
by 125-135 GeV). On a closer look, the most drastic drop occurs
for the $jets+\slashed E_T$ final state. This is explained by
referring to figure \ref{fig:branchings} where one finds that
the decay branching fraction for $Q_1 \to q B_1$ that contributes
actively to the said final state suffers by a huge margin when
$r'_Q$ goes from -0.1 to -0.5. Another feature that emerges from 
table \ref{table:predictions} is that the yields in the leptonic modes are more 
pronounced than that in the leptonically quiet mode. This can be
understood from the fact that the branching fractions of the KK 
quarks and the gluon to $W_1^\pm$ and $Z_1$ are much larger than
that to $B_1$ and that $W_1^\pm$ and $Z_1$ decay entirely into leptons and missing
particles.

\begin{table}[h]
\begin{center}
\begin{tabular}{||c||c|c|c|c|c||c|c|c||}
\hline
\hline
Scenario        &  \multicolumn{5}{|c|}{nmUED} 
                 & \multicolumn{3}{|c|}{mUED {($\Lambda R=10$)}} \\
                &  \multicolumn{5}{|c|}{$R^{-1}=1$ TeV} 
                 & \multicolumn{3}{|c|}{$R^{-1}$/spectrum in TeV} \\
\hline
$r'_Q$          & -0.1 & -0.5 & -0.7 & -1.0 & -1.5 & $R^{-1}$=1.0 &  1.4 & 1.6 \\
\hline
$m_{Q_1}$ (TeV) & 1.07 & 1.39 & 1.52 & 1.65 & 1.77 &  & &  \\
 & & & & & & $m_{G_1}$ $\approx$ 1.15 & $\approx$ 1.60  & $\approx$ 1.83 \\
& & & & & & $m_{Q_1}$ $\approx$ 1.09 & $\approx$ 1.53  & $\approx$ 1.75 \\
\hline
\hline
                              &     &     &    &    &     &      &     &    \\
          $jets+\slashed E_T$ & 466 &  27 & 15 & 10 &  83 & 1396 & 158 & 50 \\
$jets+ 1 \ell + \slashed E_T$ & 332 &  68 & 39 & 26 & 215 &  804 &  88 & 31 \\
$jets+ 2 \ell + \slashed E_T$ & 143 &  62 & 35 & 22 & 205 &  371 &  42 & 15 \\
\hline
\hline
\end{tabular}
\caption{Parton level yields (in fb) for different final states for 
varying $r'_Q$ with $r'_G=-0.7$ and $R^{-1}=1$ TeV (leading to
$m_{G_1}=1.52$ TeV) at 14 TeV LHC.  Also indicated are the 
corresponding numbers for mUED. Jets (inclusive) are comprised of 
four light flavours while the charged leptons contain only electrons 
and muons.  QCD renormalization and factorization scales are set to 
the sum of the masses of the final state particles 
(level 1 KK quarks and/or gluon) produced in the strong scattering.}
\label{table:predictions}
\end{center}
\end{table}

For the mUED part of figure \ref{table:predictions} the chosen values 
of $R^{-1}$ take care of the range of masses for KK quark/KK gluon 
that were used in the nmUED case. 
Note that the nmUED yields are computed for a fixed $m_{G_1}$ while 
$m_{Q_1}$ varies. For $m_{G_1}$ (1.52 TeV) that we employ in nmUED, a similar 
$R^{-1}$ in the two cases (1 TeV) gives larger yields for the mUED case.
The reason is simple and as follows. The mUED spectrum is dominantly 
determined by $R^{-1}$
and $R^{-1}$=1 TeV gives a much lighter ($\sim 1.15$ TeV) KK gluon in comparison
to the nmUED case in hand. 
Table \ref{table:predictions} reveals that the masses are comparable in the 
two scenarios when $r'_Q=-0.7$ in nmUED and $R^{-1}=1.4$ TeV in mUED. There also one
finds that the rates are appreciably smaller for the nmUED case the most
drastic difference being in the $jets+\slashed E_T$ final state. The reason
behind this has been discussed earlier. On the other hand, the closest
possible faking in rates occur in some of the leptonic modes for
$r'_Q=-0.7$ and $r'_Q=-1.0$ with {$R^{-1}=1.0$ TeV} in nmUED and $R^{-1}=1.6$ TeV
in mUED. However, it is crucial to note that the rates in the all jets final
state can be used as a robust discriminator between nmUED and mUED scenarios.

Thus, the pattern that exists among the yields in different final states
could already disfavour mUED. When aided by a more thorough knowledge
of their yields over the nmUED parameter space gathered through realistic
simulations, such a study would constrain the nmUED parameter space as well.
Further, crucial improvements, either in the form of exclusion or in pinning
down the region of the parameter space is possible if some of the masses
involved can be known, even if roughly. Under such a circumstance, the data
can be simultaneously confronted by SUSY scenarios and the so-called SUSY-UED
confusion could be addressed rather closely. Clearly, this is a rather
involved study and hence will be taken up in a future study.

%%%%%%%%%%%%%%%%%%%%%%%%%%%%%%%%%%%%%%%%%%%%%%%%%%%%%%%%%%%%%%%%%%%%%%%%%%%%%%%%

\section{Conclusions and Outlook}
\label{sec:outlook}

In this work we discuss the role of non-vanishing BLTs (kinetic and Yukawa) 
in the strongly interacting sector of a scenario with one flat universal extra 
dimension and their impacts on the current and future runs at the LHC.

We solve for the resulting transcendental equations for
masses numerically and discuss in detail the resulting spectra as functions
of $\rinv$ and the (scaled) brane-localized parameters, $r'_G$ and $r'_Q$.
Unlike in {mUED} where the mass spectrum is essentially dictated only by $\rinv$, 
$r'_G$ and $r'_Q$ play major roles (in conjunction with $\rinv$) in determining the
same in the nmUED scenario. This opens up the possibility
that much larger (smaller) values of $\rinv$ (which, still could result in 
lighter (heavier) KK spectra) can remain relevant at the LHC when compared
to mUED.
Nontrivial deviations from the mUED are noted 
in the strong and electroweak interaction vertices involving the level `1' quarks. 
The deviations are found to be functions of 
$r'_G$ and $r'_Q$ only. Arguably, the most nontrivial implication of 
the presence
of non-vanishing brane-localized terms is that both masses and couplings of the
KK excitations are simultaneously controlled by these free parameters and thus, 
these become correlated.
We demonstrate the same and discuss its possible implications at the
LHC and contemplated on the role it may play in extracting the fundamental parameters
of such a scenario.

We then study the basic cross sections for production of
level `1' KK gluon and KK quarks (excluding the KK top quark) 
as functions of the free parameters of the scenario
at two different LHC-energies, 8 TeV and 14 TeV. It is noted that, when compared 
to the same in mUED, for a given $\rinv$, wildly varying yields are possible. 
This is because the final state 
KK gluon and quarks can now have masses freely varying over wide ranges.
The top quark sector is kept out of the ambit of this work since the structure
and the resulting phenomenology of the same are rather involved, as usual.
On top of that, there is a new possibility of level-mixing triggered by BLTs.
Thus, this sector deserves a dedicated study. 

It is pointed out that even if the  level 1 KK gluon and the quarks 
happen to have masses compatible with mUED,
they could actually result
from an nmUED-type scenario with a value of $\rinv$ different from that in the mUED case.
Although the presence of a coupling ($g'_{G_1 Q_1 Q_0}$) with modified strength
can signal an nmUED-like scenario, such departure{s} {are} expected to be
miniscule. This is since for a given $\rinv$, an mUED-like 
spectrum is obtained only with $r'_G \simeq r'_Q$ for which 
deviations in the said coupling remain negligible.

Further, an nmUED-type scenario where the masses of the KK excitations are 
much less constrained, can fake SUSY more completely than a conventional mUED scenario.
Theoretically, one well-known approach to discriminate between these scenarios, 
is to compare the cross sections; the expectation being the same
to be larger for UED for a given set of masses in the final state 
(noting that both scenarios have the respective couplings of equal strengths).
However, it is unlikely that a SUSY-like spectrum {(unless it is degenerate)} could emerge from
an nmUED scenario of the present type without making the
deviation in coupling large from the corresponding SUSY values (which are identical 
to the corresponding SM or the mUED values). Thus, if $g'_{G_1 Q_1 Q_0}<1$, this
may bring {down an otherwise large} nmUED cross section close
to the SUSY value.

To get an idea of the actual rates for different final states (comprised of jets, 
leptons and missing energy) we computed the branching fraction of different 
excitations that appear in the cascades. For this we bring in an EW sector (with
gauge bosons and leptons and neutrinos) which resembles mUED. Some contrasting features
with respect to mUED and SUSY are noted in the form of inverted branching probabilities
to jets and leptons. This would result in an enhanced (depleted) lepton-rich (jet-rich) events
at the LHC in an nmUED-type scenario. The feature can be exploited for partial
amelioration of the infamous SUSY-UED confusion. It was also demonstrated that the latest
LHC data can rule out (conservatively) $\rinv$ up to around 1 TeV under some reasonable
set of assumptions. A rigorous framework for detailed LHC-analyses of 
such a scenario including the detector effects and complete implementation of the EW sector 
(with EW BLTs) in event generators like MadGraph is highly warranted.
This would be {the subject of a future work}.

It should be kept in mind that the nmUED scenario considered in this work is of
a rather prototype variety with some generic features {governed by three to four
basic parameters. This is a modest number for a new physics scenario. Hence, such a
scenario is much
more tractable than many of its SUSY counterparts.} Nonetheless, this
already offers a host
of rich, new effects that can be studied at the LHC. Note that the brane-localized
parameters we consider are all blind to flavours, the $SU(2)_W$ gauge 
quantum numbers and independent of the locations of the orbifold fixed-points they appear at.
Moreover, wherever appropriate, we assumed some of them ($r_Q= r_Y$ or 
$r'_Q=r'_Y$, for that matter) 
are equal. Deviations from any of these assumptions would have
important consequences.
On the other hand, in the nmUED scenario, radiative corrections to the KK-spectrum
can be expected to be somewhat significant just as they are in the case of conventional mUED. 
However, unlike in mUED where these corrections are the sole source of mass-splittings among
an otherwise degenerate set of KK excitations, the nmUED spectrum may already come with a considerable
splitting at a given KK-level even at the tree level, thus diluting the role of radiative corrections.

In any case, knowledge of the strong-production rates, {supplemented with} some crucial
information on decays of involved KK excitations, {lays the groundwork for initiating} phenomenological
studies of the nmUED scenario discussed in this work. The present work, thus, serves as a launch-pad 
to undertake a thorough analysis of such scenarios at the LHC.

%
%\vskip 10pt
\newpage
\noindent
{\bf Acknowledgments}
\vskip 7pt
\noindent
KN and SN are partially supported by funding available from the Department of 
Atomic Energy, Government of India for the Regional Centre for Accelerator-based
Particle Physics (RECAPP), Harish-Chandra Research Institute. The authors like 
to thank P. Aquino, J. Chakrabortty, N. D. Christensen, O. Mattelaer and A. Pukhov for very helpful
discussions on issues with FeynRules and CalcHEP and U. K. Dey
for many helpful discussions. AD thanks Department of Physics, University of Florida, 
Gainesville, USA for his sabbatical-stint during which this project was initiated.
The authors acknowledge the use of computational facility available at RECAPP and 
thank Joyanto Mitra for technical help.

%%%%%%%%%%%%%%%%%%%%%%%%%%%%
\appendix
\section*{Appendix}
%%%%%%%%%%%%%%%%%%%%%%%%%%%%
\section{Feynman rules}
%%%%%%%%%%%%%%%%%%%%%%%%%%%%

We write down the concrete forms of Feynman rules where we take all 4D momenta as incoming.
Details of our conventions are given in section~{\ref{sec:spectrum}.
Note that we omit the rule for the quartic coupling involving only $G_{\mu}^{(1)}$ since it is not important in LHC phenomenology.

\al{
\raisebox{-13mm}{\includegraphics[width=45mm, height=30mm, clip]{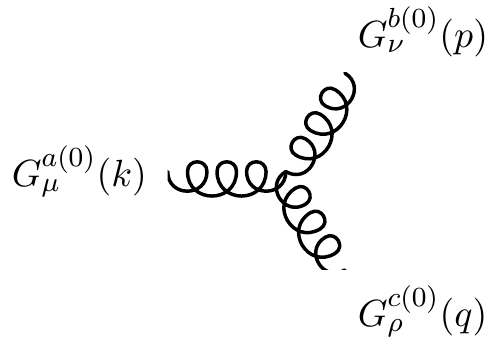}}
		&\quad= {g_{4s} f^{abc} \Big[ \eta^{\mu\nu}(k-p)^{\rho} + \eta^{\nu\rho}(p-q)^{\mu} + \eta^{\rho\mu}(q-k)^{\nu} \Big]}, \\
\raisebox{-17mm}[18mm][0mm]{\includegraphics[width=35mm, height=35mm, clip]{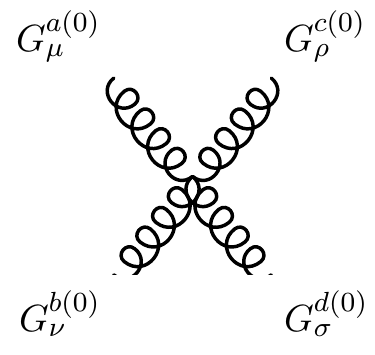}}
		&\quad= -ig_{4s}^2 \Big[ f^{abe}f^{cde} \paren{\eta^{\mu\rho} \eta^{\nu\sigma} - \eta^{\mu\sigma} \eta^{\nu\rho}} + f^{ace}f^{bde} \paren{\eta^{\mu\nu} \eta^{\rho\sigma} - \eta^{\mu\sigma} \eta^{\nu\rho}} \notag \\
		&\phantom{=-ig\,\,} + f^{ade}f^{bce} \paren{\eta^{\mu\nu} \eta^{\rho\sigma} - \eta^{\mu\rho} \eta^{\nu\sigma}} \Big], \\
\raisebox{-13mm}[23mm][0mm]{\includegraphics[width=45mm, height=30mm, clip]{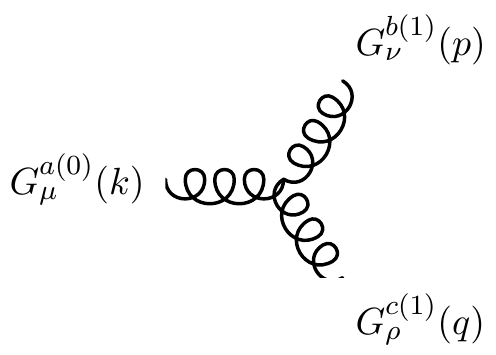}}
		&\quad= {g_{4s} f^{abc} \Big[ \eta^{\mu\nu}(k-p)^{\rho} + \eta^{\nu\rho}(p-q)^{\mu} + \eta^{\rho\mu}(q-k)^{\nu} \Big]}, \\
\raisebox{-18mm}[28mm][0mm]{\includegraphics[width=35mm, height=35mm, clip]{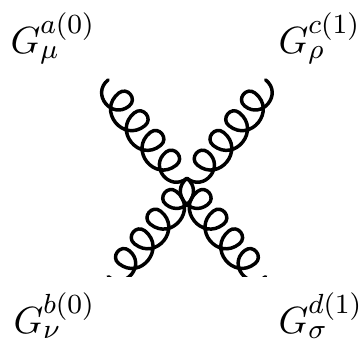}}
		&\quad= -ig_{4s}^2 \Big[ f^{abe}f^{cde} \paren{\eta^{\mu\rho} \eta^{\nu\sigma} - \eta^{\mu\sigma} \eta^{\nu\rho}} + f^{ace}f^{bde} \paren{\eta^{\mu\nu} \eta^{\rho\sigma} - \eta^{\mu\sigma} \eta^{\nu\rho}} \notag \\
		&\phantom{=-ig\,\,} + f^{ade}f^{bce} \paren{\eta^{\mu\nu} \eta^{\rho\sigma} - \eta^{\mu\rho} \eta^{\nu\sigma}} \Big],
%\raisebox{-16mm}[24mm][0mm]{\includegraphics[width=35mm, height=35mm, clip]{vertex5.pdf}}
%		&\quad=-ig_{4s}^2 (g'_{G_1G_1G_1G_1}) \Big[ f^{abe}f^{cde} \paren{\eta^{\mu\rho} \eta^{\nu\sigma} - \eta^{\mu\sigma} \eta^{\nu\rho}} \notag \\ 
%& \hspace{10mm} + f^{ace}f^{bde} \paren{\eta^{\mu\nu} \eta^{\rho\sigma} - \eta^{\mu\sigma} \eta^{\nu\rho}} + f^{ade}f^{bce} \paren{\eta^{\mu\nu} \eta^{\rho\sigma} - \eta^{\mu\rho} \eta^{\nu\sigma}} \Big], \\
}
\al{
\raisebox{-12mm}[23mm][0mm]{\includegraphics[width=30mm, bb= 0 0 108 104, clip]{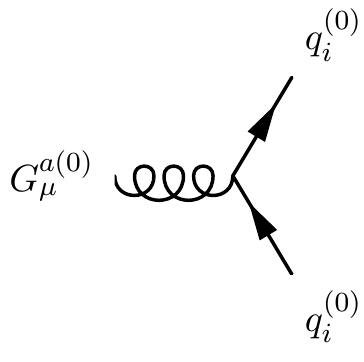}}
		&\quad= \raisebox{-12mm}{\includegraphics[width=30mm, height=30mm, clip]{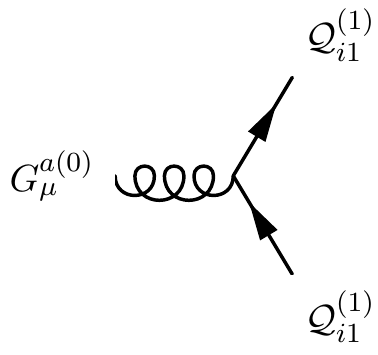}}
		\hspace{4mm}= \raisebox{-12mm}{\includegraphics[width=30mm, bb= 0 0 108 104, clip]{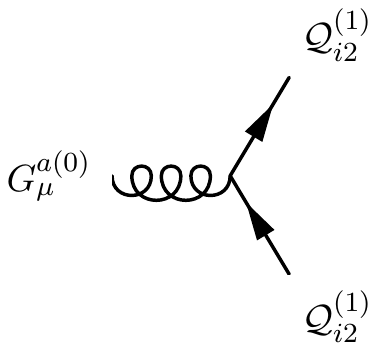}}
		\quad= { i g_{{4s}} T^{a} \gamma_{\mu}},\\
\raisebox{-12mm}[-23mm][0mm]{\includegraphics[width=30mm, bb= 0 0 108 104, clip]{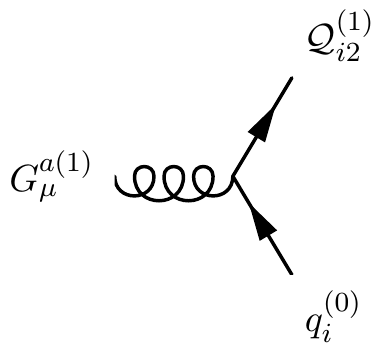}}
		&\quad= \raisebox{-12mm}{\includegraphics[width=30mm, height=30mm, clip]{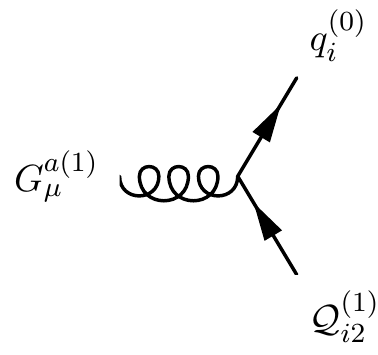}}
		\quad= { i g_{{4s}} (g'_{G_1Q_1Q_0}) T^{a} \gamma_{\mu} \paren{ v^{(1)}_{q_i R(21)} P_R + v^{(1)}_{q_i L(11)} P_L } },\\
\raisebox{-12mm}[-23mm][0mm]{\includegraphics[width=30mm, height=30mm, clip]{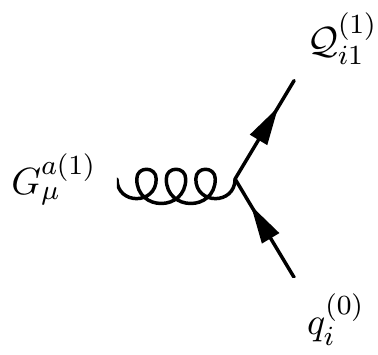}}
		&\quad= \raisebox{-12mm}{\includegraphics[width=30mm, height=30mm, clip]{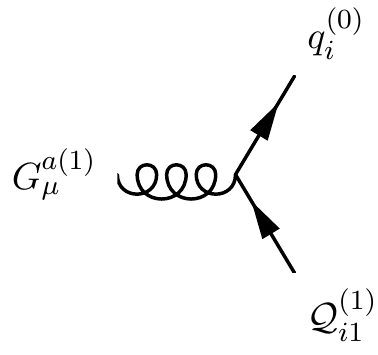}}
		\quad= { i g_{{4s}} (g'_{G_1Q_1Q_0}) T^{a} \gamma_{\mu} \paren{ v^{(1)}_{q_i R(22)} P_R + v^{(1)}_{q_i L(12)} P_L } }.
}

%
%%%%%%%%%% Bibliography %%%%%%%%%%
%

%
\end{document}